\documentclass[reprint,superscriptaddress,amsmath,amssymb,prd,nofootinbib,longbibliography]{revtex4-2}

\usepackage{amssymb,amsmath,epsfig}
\usepackage{mathrsfs}
\usepackage{mathrsfs}
\usepackage[usenames,dvipsnames]{color}
\usepackage[bookmarks]{hyperref}
\usepackage[svgnames]{xcolor}
\usepackage{subfigure}
\usepackage{times}
\usepackage{yhmath}
\usepackage{braket}
\usepackage{soul}

\newcommand{\Real}{\mathbb{R}}

\newcommand{\re}{\mbox{Re}}

\newcommand{\proof}{\noindent {\bf Proof. }}

\newcommand{\qed}{\hfill \fbox{} \vspace{.3cm}}

\newtheorem{definition}{Definition}
\newtheorem{lemma}{Lemma}

\newtheorem{theorem}{Theorem}

\begin{document}

\title{Relativistic dissipative fluids in the trace-fixed particle frame: Strongly hyperbolic quasilinear first-order evolution equations}

\author{J. F\'elix Salazar}
\address{Departamento de Matem\'aticas Aplicadas y Sistemas, Universidad Aut\'onoma Metropolitana-Cuajimalpa, 05348 Cuajimalpa de Morelos, Ciudad de M\'exico M\'exico.}

\author{Ana Laura Garc\'ia-Perciante}
\address{Departamento de Matem\'aticas Aplicadas y Sistemas, Universidad Aut\'onoma Metropolitana-Cuajimalpa, 05348 Cuajimalpa de Morelos, Ciudad de M\'exico M\'exico.}

\author{Olivier Sarbach}
\address{Departamento de Matem\'aticas Aplicadas y Sistemas, Universidad Aut\'onoma Metropolitana-Cuajimalpa, 05348 Cuajimalpa de Morelos, Ciudad de M\'exico M\'exico.}
\address{Instituto de F\'isica y Matem\'aticas,
Universidad Michoacana de San Nicol\'as de Hidalgo,
Edificio C-3, Ciudad Universitaria, 58040 Morelia, Michoac\'an, M\'exico.}

\begin{abstract}
In this paper we derive a new first-order theory of relativistic dissipative fluids by adopting the trace-fixed particle frame. Whereas in a companion letter we show that this theory is hyperbolic, causal and stable at global equilibrium states, here we prove that the full nonlinear system of equations can be cast into a first-order quasilinear system which is strongly hyperbolic. By rewriting the system in first-order form, auxiliary constraints are introduced. However, we show that these constraints propagate, and thus our theory leads to a well-posed Cauchy problem.
\end{abstract}

\date{\today}

\pacs{04.20.-q,04.40.-g}

\maketitle

\section{Introduction}
\label{Sec:Introduction}

A persistent challenge in theoretical physics is the formulation of a theory of dissipative fluids which aligns with both thermodynamics and (special or general) relativity. The need for obtaining such a formulation is reinforced by recent experiments and observations in high-energy physics and relativistic astrophysics. A prominent example is the understanding of the dynamics and transport properties of the quark-gluon plasma produced in heavy-ion colliders~\cite{DKK2016,DH2020,Shen2020}. Moreover, the interpretation of data obtained from detections of gravitational radiation generated by binary neutron star \cite{LIGO2017_A,LIGO2017_B} or black hole-neutron star mergers \cite{LIGO2021}, requires an accurate description of viscous fluids in the strong gravity regime~\cite{ChR2021,Hatton2024}. Also, it has been found that dissipative effects can significantly alter the properties of the outflow and electromagnetic signals in accretion processes~\cite{ChR2021}.

A relativistic theory of dissipative fluids which is suitable for addressing the aforementioned problems should satisfy a minimal set of requirements. On the one hand, this entails that the evolution equations give rise to a well-posed Cauchy problem. Furthermore, this evolution system should be causal, meaning that no physical modes propagate faster than the speed of light. Local in time well posedness and finite speed of propagation can be conveniently achieved by requiring that the evolution equations are strongly hyperbolic. On the other hand, mandatory requisites of the theory are the fulfillment of the second law of thermodynamics and the stability of equilibrium states, in the sense that small perturbations of global equilibrium configurations in Minkowski space decay in time. Satisfying all these criteria has proven to be a challenging task and has led to a plethora of proposals since the early approaches by Eckart and Landau-Lifshitz (see Refs.~\cite{Van2020,SZ2020,Gavassino2021,RDNR2024} for recent reviews).

Recently, a new formulation known in the literature as Bemfica-Disconzi-Noronha-Kovtun (BDNK) theory~\cite{BDN2018,BDN2019,pK19,Hoult2020,Bemfica2022,Kovtun2022,Rocha2022,Disconzi2024} has attracted a lot of attention. This theory is based on a gradient expansion in the state variables $(n,\varepsilon,u^\mu)$ (representing, respectively, the particle density, energy density, and four-velocity of the fluid when in equilibrium), or equivalent variables involving the chemical potential. When truncated to first order, and under suitable restrictions on the transport coefficients, the theory can be shown to satisfy the aforementioned requirements in the following sense: In~\cite{Bemfica2022}, a strongly hyperbolic and causal system is obtained which is shown to give rise to non-negative entropy production (within the limits of validity of the theory) and  to stable global equilibria, provided the transport coefficients obey an ample set of inequalities. Also, Kovtun et. al.~\cite{pK19,Hoult2020,Kovtun2022} obtain dissipative fluid theories which are causal, stable, and shown to be weakly hyperbolic. A key observation in BDNK theory is to relax the restriction of the Eckart or Landau-Lifshitz frames which are usually adopted and result in theories that are not only unstable but are not even hyperbolic~\cite{whlL85}. The frame in~\cite{Bemfica2022} is partially determined by fixing $n$ and $u^\mu$ to those matching the current density $J^\mu$, i.e. such that $J^\mu = n u^\mu$; however, $\varepsilon$ is not required to match the energy density measured by an observer comoving with $u^\mu$, as in the Eckart frame. Instead, the stress-energy tensor is determined by imposing constitutive relations with parameters that must satisfy certain inequalities that enforce the desired properties of the system. Similarly, the work in~\cite{pK19,Hoult2020} does not impose specific matching conditions and determines the frame through suitable constitutive relations. For BDNK hydrodynamics arising from the fluid-gravity correspondence; see, for instance, Refs.~\cite{Kovtun2022,lClL23}.

In this article, we adopt an approach which, although similar in spirit to BDNK theory, imposes matching conditions which completely fix the frame. As in~\cite{Bemfica2022}, $n$ and $u^\mu$ are fixed by the current density. However, in contrast to that work, we fix $\varepsilon$ through the trace of the stress-energy tensor $T^{\mu \nu}$. More precisely, we work with state variables $(n,T,u^\mu)$ (with $T$ representing the temperature when in equilibrium) which parametrize the equilibrium configuration that matches the expressions for $J^\mu$ and $T^{\mu}{}_{\mu}$. This choice, which we shall refer to as the \emph{trace-fixed particle frame} (or TFP frame), can be motivated from kinetic theory by noticing that it fixes the first few moments of the distribution function \cite{Stewart1971}. Further, it is worthwhile to point out that this particular frame is well adapted to conformal fluids for which $T^{\mu}{}_{\mu}=0$.

Apart from the three transport coefficients $\eta$, $\zeta$, and $\kappa$, denoting the shear and bulk viscosities and the thermal conductivity, respectively, our theory possesses two additional coefficients, $\Gamma_1$ and $\Gamma_2$. However, they do not affect the physical content of the theory up to and including terms which are first order in the gradients of the state variables. In a companion letter~\cite{aGjSoS2024a} we show that a suitable choice for $\Gamma_2$ leads to a system of equations which is strongly hyperbolic and causal when linearized around an homogenous equilibrium state, provided the following inequality is fulfilled (with $k_B$ denoting Boltzmann's constant):
\begin{equation}
1 + 2\frac{\eta}{\kappa} 
 \leq \frac{e}{k_B T}.
\label{Eq:FinalHypoCausalBound}
\end{equation}
Notice that this condition only involves the temperature parameter $T$, the internal energy per particle $e$, and the transport coefficients $\eta$ and $\kappa$. Further, it was verified in~\cite{aGjSoS2024a} that this inequality is satisfied for a simple gas in two or three space dimensions with a hard disk or sphere cross section. Additionally, equilibrium states for such gases are proven to be stable under the assumption that $\Gamma_1$ is sufficiently large.

The main goal of this article is to show that, given the validity of the inequality~(\ref{Eq:FinalHypoCausalBound}), our theory is governed by evolution equations which are strongly hyperbolic and causal in the full nonlinear regime. This leaves the single free coefficient $\Gamma_1$ which is only restricted by the stability condition discussed in the companion letter. Moreover, as we will show in separate work~\cite{cGaGoS25}, our theory emerges naturally from the Chapman-Enskog approach in relativistic kinetic theory.

Our theory leads to a system of equations for the  fields $(n,T,u^\mu,\epsilon,\mathcal{Q}^\mu)$, where $\epsilon$ denotes the off-equilibrium contribution to the energy density and $\mathcal{Q}^\mu$ the heat flow. These equations provide an explicit relation between the first-order derivatives of these fields along the velocity vector $u^\mu$ and their derivatives $D_\mu$ in directions orthogonal to $u^\mu$. In particular, this relation involves \emph{second-order} derivatives of the velocity field with respect to $D_\mu$ originating from the divergence of the shear. Hence, one obtains an evolution-type system which is mixed first and second order which could in principle be addressed using the  ``principle of frozen coefficients"~\cite{Kreiss89} and a pseudodifferential reduction (see for example Refs.~\cite{gNoOoR04,oSmT12} and references therein). Whereas in our companion letter~\cite{aGjSoS2024a} we show that the frozen coefficient problems at global equilibrium states are well posed when the condition~(\ref{Eq:FinalHypoCausalBound}) is satisfied, a technical difficulty arises when considering the nonlinear problem. This is due to the fact that the velocity field is not hypersurface orthogonal in general, and thus when introducing local coordinates and writing the system as partial differential equations (PDEs), second order time derivatives of the velocity field appear.

To circumvent this issue we rewrite the equations as a first-order system at the cost of introducing auxiliary constraints. This is achieved by introducing the directional derivatives $D_\mu$ of the temperature, particle density and the velocity field as independent variables and deriving suitable evolution equations for them. In this way one obtains a first-order quasilinear evolution system for the state variables $(n,T,u^\mu)$, their  derivatives along $D_\mu$, and the fields $\epsilon$ and $\mathcal{Q}^\mu$. In particular, this system includes evolution equations for the expansion, shear and vorticity associated with $u^\mu$ which play an important role in the study of congruences of timelike curves~\cite{HawkingEllis-Book,Wald1984,Ellis2012}.

In principle, the hyperbolicity of the resulting constrained system can be analyzed based on the methods developed in~\cite{fAoR20,fA22,fAoRdH24}. In this article, by exploring the freedom of adding the constraints to the evolution equations, we provide an explicit set of evolution equations which is proven to be strongly hyperbolic. An essential tool in this proof is the covariant (coordinate-independent) formalism described in Refs.~\cite{rG96,oR04}, which allows one to analyze directly the principal symbol associated with the directional derivatives $D_\mu$ whose structure is much simpler than the one obtained from the PDE problem.

Finally, to complete the proof that our fluid equations lead to a well-posed Cauchy problem, one needs to show that the constraints propagate. This means that one needs to prove that initial data of the first-order system which satisfy the constraints lead to solutions for which the constraints are everywhere satisfied.

In order to achieve these goals, we start in Sec.~\ref{Sec:FluidEquations} by deriving the constitutive equations which, together with the balance equations, describe our theory. For this, we perform a change of frame from Eckart to TFP and contrast the resulting constitutive relations with the ones typically chosen in BDNK theory. In Sec.~\ref{Sec:Results} we establish the evolution equations that are analyzed in this work, summarize the important properties of the linearized system discussed in~\cite{aGjSoS2024a}, and then state the local well-posedness result for the full nonlinear system in Theorem~1. Sections~\ref{Sec:FOQ}--\ref{Sec:ConstraintPropagation} are devoted to the proof of this theorem. In Sec.~\ref{Sec:FOQ} we enlarge the evolution system by introducing new variables in order to rewrite it as a first-order quasilinear system of equations with constraints. This system is shown to be strongly hyperbolic and causal in Sec.~\ref{Sec:Hypo}, and in Sec.~\ref{Sec:ConstraintPropagation} we prove that the constraints propagate, implying that our theory is described by a well-posed Cauchy problem. Conclusions are drawn in Sec.~\ref{Sec:Conclusions}. Some technical, yet important details are included in Appendices~\ref{App:FirstOrderDissipativeFluids}--\ref{App:StronglyHyperbolicPDE}. In App.~\ref{App:FirstOrderDissipativeFluids} we provide a summary of the first-order frame invariant transport parameters introduced by Kovtun~\cite{pK19} and generalize them to the presence of an external electromagnetic field. In App.~\ref{App:Entropy} we recall the definition of the entropy current and show that our theory leads to a non-negative entropy production within its domain of validity. Several commutator identities which are used to derive our first-order system are summarized in  App.~\ref{App:Commutators}. Finally, a short summary of the theory developed in~\cite{rG96,oR04} and its application to our problem are given in App.~\ref{App:StronglyHyperbolicPDE}.

We work on a $(d+1)$-dimensional smooth ($C^\infty$) background spacetime $(M,g)$ which is assumed to be globally hyperbolic and time oriented and whose metric has signature $(-,+,\ldots,+)$, and we adopt units in which the speed of light is one. Greek indices run over $0,1,\ldots,d$ and $\Delta_\mu{}^\nu := \delta_\mu{}^\nu + u_\mu u^\nu$ denotes the operator that projects orthogonal to the velocity vector $u^\mu$. The $2$-form $F = \frac{1}{2} F_{\mu\nu} dx^\mu\wedge dx^\nu$ represents a given, smooth $(C^\infty)$ electromagnetic field on $(M,g)$. The brackets $(\ )$, $\langle \ \rangle$ and $[\ ]$ denote the symmetric, the symmetric trace-free and the antisymmetric parts of a tensor field, respectively.

\section{Hydrodynamic equations in the trace-fixed particle frame}
\label{Sec:FluidEquations}

In this section, the transport equations for a charged fluid in the TFP frame are derived. For this purpose, we begin by stating the balance equations for the particle current density $J^{\mu}$ and stress-energy tensor $T^{\mu\nu}$ namely,
\begin{equation}
\nabla_{\mu}J^{\mu}=0\quad\text{and}\quad
\nabla_{\mu} T^{\mu\nu} + q J_\mu F^{\mu\nu} = 0,
\label{eq:euler}
\end{equation}
where $\nabla_{\mu}$ stands for the covariant derivative associated with the metric and $q$ is the charge of the particles.

\subsection{Equilibrium equations}

In equilibrium, or for a perfect fluid, one has
\begin{equation}
J^{\mu}=nu^{\mu},\quad T^{\mu\nu}=n e u^{\mu}u^{\nu}+p\Delta^{\mu\nu},
\label{eq:II.3}
\end{equation}
where $n$ is the particle number density, $e$ the internal
energy per particle and $p=nk_{B}T$ the hydrostatic pressure,
with $T$ being the temperature. The balance equations in this case correspond to the relativistic
Euler equations in the presence of an external electromagnetic field which read: 
\begin{align}
 & \dot{n}+n\theta=0,
\label{eq:II.4}\\
 & a_{\mu}+\frac{1}{nh}D_{\mu}p -\frac{q}{h}E_{\mu}=0,
\label{eq:II.5}\\
 & \frac{\dot{T}}{T}+\frac{k_{B}}{c_{v}}\theta=0,
\label{eq:II.6}
\end{align}
where here and in the following, a dot refers to the derivative with respect to the comoving observer's proper time, i.e. $\dot{n} = u^\mu\nabla_\mu n$ and similarly for $\dot{T}$, and $D_\mu p := \Delta_{\mu}{}^\nu\nabla_\nu p$ (see Appendix~\ref{App:Commutators} for the generalization of these operators to tensor fields orthogonal to $u^\mu$). Furthermore, $\dot{u}_{\mu}:=a_{\mu}:= u^\nu\nabla_\nu u_\mu$ and $\theta:=\nabla_{\mu}u^{\mu}$
stand for the acceleration and the expansion, respectively. The specific heat at constant volume $c_{v}$ and the enthalpy $h$ (both quantities
measured per particle) are defined as $c_{v}:=\left. \frac{\partial e}{\partial T}\right|_{n}$
and $h:=e + k_{B}T$, and $E^{\mu}:=u_{\nu}F^{\mu\nu}$
stands for the electric field measured by a comoving observer.

\subsection{Constitutive relations}

Out of equilibrium one can in general write
\begin{eqnarray}
J^{\mu} & =& \mathcal{N}u^{\mu}+\mathcal{J}^{\mu},
\label{eq:II.6} \\  
T^{\mu\nu} & =&\mathcal{E}u^{\mu}u^{\nu}+\mathcal{P}\Delta^{\mu\nu}+2 u^{(\mu}\mathcal{Q}^{\nu)}+\mathcal{T}^{\mu\nu},
\label{eq:II.8}
\end{eqnarray}
where $\mathcal{J^\mu}$, $\mathcal{Q^\mu}$ and $\mathcal{T}^{\mu\nu}$ are orthogonal to $u^\mu$ and $\mathcal{T}^{\mu\nu}$ is symmetric and trace-free. The quantities $\mathcal{N}$, $\mathcal{E}$, $\mathcal{P}$, $\mathcal{J}^{\mu}$, $\mathcal{Q}^\mu$, $\mathcal{T}^{\mu\nu}$ are expanded as their equilibrium part plus a nonequilibrium correction expressed as a linear combination of first-order derivatives of the state variables $(n,\,T,\,u^\alpha)$ and $E^\mu$, see Eqs.~(\ref{Eq:ConstitutiveN}-\ref{Eq:ConstitutiveT}) in Appendix~\ref{App:FirstOrderDissipativeFluids}.\footnote{We assume that $E^\mu$ scales like the first-order gradients of the state variables, which is consistent with Eq.~(\ref{eq:II.5}).}  Notice that, whereas in an equilibrium situation the parameters contained in the decompositions for $J^\mu$ and $T^{\mu\nu}$ given in Eq.~(\ref{eq:II.3}) can be clearly and uniquely associated with the state variables $n$, $u^\mu$ and $T$, \footnote{Note that the temperature $T$ is uniquely determined by $e$ as long as $c_v=\left(\frac{\partial e}{\partial T}\right)_n>0$.} out of equilibrium this is not the case. The freedom of assigning the state variables $(n,\,T,\,u^\mu)$ to $J^\mu$ and $T^{\mu\nu}$ for an out of equilibrium state is known as a {\it choice of frame}. Different frames may lead to theories with different features regarding hyperbolicity, causality and stability. In the following, we single out the TFP frame and show that it leads to a physically sound theory.
This particular frame is characterized by setting $\mathcal{N}=n$, $\mathcal{J}^\mu=0$ and $-\mathcal{E}+d \mathcal{P}=-ne+dp$ in Eqs. (\ref{eq:II.6}, \ref{eq:II.8}), which corresponds to fixing $J^\mu$ and $T^\mu{}_\mu$ to their equilibrium values. In order to uniquely determine the temperature from the trace, the condition $\left[\frac{\partial}{\partial T}(-n e + dp)\right]_n > 0$ is required, which is equivalent to $c_v < d k_B$ and will be assumed in what follows.

The system of constitutive equations is established by starting from the well-known relations in Eckart's frame and using the tools described in Appendix \ref{App:FirstOrderDissipativeFluids} to transform them to the TFP frame. In the Eckart frame \cite{EckartIII}, one considers $\nu_i=\varepsilon_i=\gamma_i=0$ in Eqs. (\ref{Eq:ConstitutiveN}-\ref{Eq:ConstitutiveT}), such that
\begin{align}
J^{\mu} & =nu^{\mu},\\
T^{\mu\nu} & =neu^{\mu}u^{\nu}+\mathcal{P}\Delta^{\mu\nu}+2 u^{(\mu}\mathcal{Q}^{\nu)}+\mathcal{T}^{\mu\nu},
\end{align}
where
\begin{align}
\mathcal{P} & =p-\zeta\theta,\label{PEckart}\\
\mathcal{Q}^{\mu} & =-\kappa\left(\frac{D^{\mu}T}{T}+\dot{u}^{\mu}\right),\label{QEckart}\\
\mathcal{T}^{\mu\nu}&=-2\eta\sigma^{\mu\nu},\label{TauEckart}
\end{align}
with $\kappa$, $\zeta$, $\eta$ being the thermal conductivity and the bulk and shear viscosities respectively, here assumed to be functions of the temperature $T$ only.\footnote{For instance, this assumption is valid for a simple kinetic gas with a hard disk or sphere cross section, see Appendix~A of Ref.~\cite{aGjSoS2024a} and references therein.} Further,  $\sigma^{\mu\nu}$ denotes the traceless symmetric part of the spatial gradient of $u^\mu$, that is,
$\sigma^{\mu\nu}:=\Delta^{\langle \mu}{}_\alpha \Delta^{\nu \rangle}{}_\beta\nabla^\alpha u^\beta$.
Next, we observe that the transport equations in the TFP frame are described by setting $\nu_i=\gamma_i=\varepsilon_i-d\pi_i=0$ in Eqs. (\ref{Eq:ConstitutiveN}-\ref{Eq:ConstitutiveT}). According to the work by Kovtun~\cite{pK19}, which is summarized and generalized to include the electromagnetic field in Appendix~\ref{App:FirstOrderDissipativeFluids}, the following eight quantities remain invariant when passing from the Eckart to the TFP frame: $\eta$ and
\begin{align}
f_{i}&=-\frac{k_B}{c_{v}}\varepsilon_{i} + \pi_{i},\qquad i=1,2,3,\\
\ell_i&=-\frac{1}{h}\kappa_i,\qquad i=1, 2, 3, 4,
\end{align}
as follows from Eqs.~(\ref{Eq:FrameInvariantf}, \ref{Eq:FrameInvariantl}) with $\gamma_i=\nu_i=0$ which holds in both frames. Thus, the constitutive equation for the heat flux and $\mathcal{T}^{\mu\nu}$ remain unaltered due to the invariance of $\ell_{i}$ and $\eta$. Taking into account that in Eckart's frame 
$f_{1}=f_{2}=0$ and $f_{3}=-\zeta$, the invariance of $f_{i}$ implies that in the TFP frame, $\varepsilon_{i}=\pi_{i}=0$ for $i=1,2$, and
\begin{equation}
\varepsilon_3= \frac{\zeta}{\frac{k_B}{c_{v}}-\frac{1}{d}},\qquad
\pi_3 = \frac{1}{d}\varepsilon_3.
\end{equation}
In view of these results, one can write
\begin{equation}
    \mathcal{E} = ne + \epsilon, \quad \mathcal{P} = p + \frac{\epsilon}{d},
\end{equation}
such that the particle current and the stress-energy tensor in the TFP frame are given by
\begin{align}
J^{\mu}&=nu^{\mu},
\label{JTFP} \\T^{\mu\nu}&=\left(ne+\epsilon\right)u^{\mu}u^{\nu}+\left(p+\frac{\epsilon}{d}\right)\Delta^{\mu\nu}+2u^{(\mu}\mathcal{Q}^{\nu)}+\mathcal{T}^{\mu\nu},
\label{TTFP}
\end{align}
where the constitutive equations are
\begin{equation}
\epsilon = \frac{\zeta\theta}{\frac{k_B}{c_{v}}-\frac{1}{d}}, 
\end{equation}
and Eqs.~(\ref{QEckart},\ref{TauEckart}).

The next observation which is crucial to obtain a hyperbolic system is the realization that one has the freedom to add the following combinations of terms:
\begin{equation}
 \frac{\dot{T}}{T}+\frac{k_B}{c_{v}}\theta,
 \qquad
 a_\mu + \frac{D_\mu p}{nh} - \frac{q}{h} E_\mu,
\label{Eq:CombinationsEuler}
\end{equation}
which are second order in the gradients (see Eqs.~(\ref{Eq:EulerTrunc2}) and (\ref{Eq:EulerTrunc3})), to the right-hand side of the constitutive equations.\footnote{In the TFP frame the right-hand side of Eq.~(\ref{Eq:EulerTrunc1}) is exactly zero and hence nothing is gained by adding this equation to the right-hand side of the constitutive relations.} Specifically, we add the first combination to $\epsilon$ with the factor $\zeta\left(\frac{k_B}{c_{v}}-\frac{1}{d}\right)^{-2}(\Gamma_1-1)$ and the second one to $\mathcal{Q}^\mu$ with a factor $\kappa\Gamma_2$, with $\Gamma_i$ arbitrary dimensionless coefficients which may depend on $T$. This leads to the final form of the constitutive relations
\begin{align}
\epsilon&=-\frac{\zeta}{\left(\frac{k_B}{c_{v}}-\frac{1}{d}\right)^{2}}\left[ \frac{\dot{T}}{T} + \frac{\theta}{d} - \Gamma_1\left( \frac{\dot{T}}{T} + \frac{k_B}{c_v}\theta \right) \right],
\label{Eq:ConstitutiveRelations_e}\\
\mathcal{Q}_\mu&=-\kappa \left[\frac{D_\mu T}{T} + a_\mu - \Gamma_2\left( a_\mu + \frac{D_\mu p}{nh} - \frac{q}{h} E_\mu\right) \right],
\label{Eq:ConstitutiveRelations_q}\\
\mathcal{T}_{\mu\nu}&=-2\eta\sigma_{\mu\nu}.
\label{Eq:ConstitutiveRelations_Pi}
\end{align}
Note that variations of the coefficients $\Gamma_1$ and $\Gamma_2$ correspond to changes of representation as defined in~\cite{JNET24} and Appendix~\ref{App:FirstOrderDissipativeFluids}. However, given the parametrization of the particle current density and stress-energy tensor in Eqs.~(\ref{JTFP},\ref{TTFP}), such variations exactly maintain the TFP frame.

At this point, we would also like to stress that our constitutive relations~(\ref{Eq:ConstitutiveRelations_e}--\ref{Eq:ConstitutiveRelations_Pi}) are \emph{different} from the ones postulated in Eqs.~(11) of Ref.~\cite{Bemfica2022}. Whereas both formulations are based on a particle frame in which $J^\mu = n u^\mu$, our choice of fixing the trace of the stress-energy tensor would imply that
\begin{equation}
0 = (3\tau_P - \tau_\varepsilon)n\dot{e} 
 + \left[(3\tau_P - \tau_\varepsilon)p - 3\zeta) \right]\theta,
\end{equation}
for their coefficients $\tau_\varepsilon$ and $\tau_P$ and the bulk viscosity $\zeta$, which can only be satisfied if $\zeta$ vanishes identically. This is due to the fact that Ref.~\cite{Bemfica2022} adds the combinations~(\ref{Eq:CombinationsEuler}) directly to the constitutive relations in the Eckart frame. In contrast, we first perform a change of frame before adding these combinations. Moreover, when adding the combinations~(\ref{Eq:CombinationsEuler}), we do it in such a fashion to preserve the TFP frame which leaves us with the two free coefficients $\Gamma_1$ and $\Gamma_2$ instead of the three coefficients $\tau_\varepsilon$, $\tau_P$ and $\tau_Q$ in~\cite{Bemfica2022}. As we show here and in the companion letter~\cite{aGjSoS2024a}, it is much simpler to satisfy the conditions for hyperbolicity, causality and stability in the TFP frame than in the frame considered in~\cite{Bemfica2022}. Furthermore, the examples of causal theories provided by Kovtun (see, for instance, App. A in Ref. \cite{Kovtun2022}) seem to require a nontrivial first order contribution to the particle current and thus correspond to frames different than the TFP one.

Note also that the choice $\Gamma_1 = \Gamma_2 = 1$ in Eqs.~(\ref{Eq:ConstitutiveRelations_e},\ref{Eq:ConstitutiveRelations_q}) eliminates the time derivatives of $T$ and $u^\mu$ in these relations, which would lead to the usual couplings of dissipative fluxes with purely spatial gradients of the state variables. These kinds of relations are the ones found in nonrelativistic fluids and can also be obtained from relativistic kinetic theory based on the traditional Chapman-Enskog method~\cite{wI63}. However, this choice does not yield a hyperbolic system.

\section{Evolution equations and main results}
\label{Sec:Results}

The system of transport equations describing the fluid in the TFP frame is given by Eqs.~(\ref{eq:euler}) together with (\ref{JTFP}), (\ref{TTFP}), coupled to the constitutive relations obtained in the previous subsection, i.e. Eqs.~(\ref{Eq:ConstitutiveRelations_e})-(\ref{Eq:ConstitutiveRelations_Pi}). The usual procedure consists in substituting the latter in Eqs.~(\ref{eq:euler}) to obtain an evolution system for the state variables $\left(n,\,T,\,u^\mu\right)$. This leads to a set of equations that involves second-order time derivatives of $T$ and $u^\mu$. However, a system which is first order in time derivatives can be naturally obtained in terms of the variables  $\left(n,\,T,\,u^\mu,\,\epsilon,\,\mathcal{Q}^\mu\right)$.

To achieve this we assume that $\Gamma_1$ and $\Gamma_2$ are different from one, which allows us to rewrite Eqs.~(\ref{Eq:ConstitutiveRelations_e}) and (\ref{Eq:ConstitutiveRelations_q}) in the form
\begin{align}
(1 - \Gamma_1)\frac{\dot{T}}{T} &= -\left( \frac{1}{d} - \Gamma_1\frac{k_B}{c_v} \right)\theta - \left(\frac{k_B}{c_{v}}-\frac{1}{d}\right)^{2}\frac{\epsilon}{\zeta},
\label{Eq:Tdot}\\
(1 - \Gamma_2) a_\mu &= \frac{\Gamma_2 k_B T}{h}\frac{D_\mu n}{n} - \left( 1 - \frac{\Gamma_2 k_B T}{h} \right)\frac{D_\mu T}{T}
\nonumber\\
 &- \Gamma_2\frac{q}{h} E_\mu 
 - \frac{1}{\kappa}\mathcal{Q}_\mu,
\label{Eq:udot}
\end{align}
and use them as evolution equations for $T$ and $u^\mu$. The equations for the remaining variables $(n,\epsilon,\mathcal{Q}^\mu)$ are obtained from Eqs.~(\ref{eq:euler}), where time derivatives of $T$ and $u^\mu$ are eliminated by means of Eqs.~(\ref{Eq:Tdot}, \ref{Eq:udot}). This yields the system
\begin{align}
\frac{\dot{n}}{n} &= -\theta,
\label{Eq:ndot}\\
\frac{\dot{T}}{T} &= \alpha_4\theta + \alpha_5\epsilon,
\label{Eq:TdotBis}\\
\dot{u}^\mu &= \beta_1\frac{D^\mu n}{n} + \beta_2\frac{D^\mu T}{T} + \beta_3\mathcal{Q}^\mu -\frac{q \beta_{1}}{k_{B}T}E^{\mu},
\label{Eq:udotBis}\\
\dot{\epsilon} &= -D_\mu\mathcal{Q}^\mu
 + f_\epsilon\left( n,T,\epsilon,\mathcal{Q}^\mu,\theta,\sigma_{\mu\nu},\frac{D_\mu n}{n},\frac{D_\mu T}{T} \right),
\label{Eq:Edot}\\
\dot{\mathcal{Q}}^\mu &=
2\eta D_\nu\sigma^{\mu\nu} - \frac{1}{d} D^\mu\epsilon\nonumber\\
 &+ f_\mathcal{Q}^\mu\left( n,T,\epsilon,\mathcal{Q}^\mu,\theta,\sigma_{\mu\nu},\omega_{\mu\nu},\frac{D_\mu n}{n},\frac{D_\mu T}{T} \right),
\label{Eq:qdot}
\end{align}
where we have introduced the coefficients
\begin{align}
& \alpha_4 := -\frac{\Gamma_1\frac{k_B}{c_v} - \frac{1}{d}}{\Gamma_1 - 1},
\quad
\alpha_5 := \frac{1}{\Gamma_1 - 1}\left(\frac{k_B}{c_{v}}-\frac{1}{d}\right)^{2}\frac{1}{\zeta},
\label{Eq:alphaCoeff}\\
& \beta_1 := -\frac{\Gamma_2\frac{k_B T}{h}}{\Gamma_2 - 1},
\quad
\beta_2 := \frac{1 - \Gamma_2\frac{k_B T}{h}}{\Gamma_2 - 1},
\quad
\beta_3 := \frac{1}{\Gamma_2 - 1}\frac{1}{\kappa},
\label{Eq:betaCoeff}
\end{align}
and the nonlinear terms
\begin{widetext}
\begin{align}
f_\epsilon\left( n,T,\epsilon,\mathcal{Q}^\mu,\theta,\sigma_{\mu\nu},\frac{D_\mu n}{n},\frac{D_\mu T}{T} \right) &:=
 -\frac{d+1}{d}\theta\epsilon + 2\eta\sigma^{\mu\nu}\sigma_{\mu\nu} - 2a_\mu\mathcal{Q}^\mu 
+ \frac{n T c_v}{\Gamma_1 - 1}\left( \frac{k_B}{c_v} - \frac{1}{d} \right)\left[\theta - \left(\frac{k_B}{c_{v}}-\frac{1}{d}\right)\frac{\epsilon}{\zeta} \right],
\\
f_\mathcal{Q}^\mu\left( n,T,\epsilon,\mathcal{Q}^\mu,\theta,\sigma_{\mu\nu},\omega_{\mu\nu},\frac{D_\mu n}{n},\frac{D_\mu T}{T} \right) &:=2\sigma^{\mu\nu} \left( T\frac{\partial\eta}{\partial T}\frac{D_\nu T}{T}+a_\nu\eta\right) - \frac{d+1}{d} a^\mu \epsilon
 - \frac{d+1}{d}\theta\mathcal{Q}^\mu - (\sigma^{\mu\nu} - \omega^{\mu\nu}) \mathcal{Q}_\nu
\nonumber\\
 &+ \frac{n h}{\Gamma_2 - 1}\left[ \frac{k_B T}{h}\frac{D^\mu n}{n} - \frac{e}{h}\frac{D^\mu T}{T} - \frac{q}{h} E^\mu - \frac{1}{\kappa}\mathcal{Q}^\mu \right].
\end{align}
Here, $\omega_{\mu\nu} := \Delta_{[\mu}{}^\alpha\Delta_{\nu]}{}^\beta\nabla_\alpha u_\beta$ denotes the vorticity and it is understood that $a^\mu$ is replaced by the right-hand side of Eq.~(\ref{Eq:udotBis}).
\end{widetext}

Equations~(\ref{Eq:ndot}--\ref{Eq:qdot}) constitute an evolution system for the fields $(n,T,u^\mu,\epsilon,\mathcal{Q}^\mu)$ which is first order in time and mixed first and second order in space (second-order spatial derivatives of the velocity field $u^\mu$ appear on the right-hand side of Eq.~(\ref{Eq:qdot})). In the companion letter~\cite{aGjSoS2024a} we analyzed the system when linearized about a global equilibrium configuration in Minkowski spacetime and proved the following properties:
\begin{enumerate}
\item[(a)] It is strongly hyperbolic and causal provided $T$, $e$, $\eta$, $\kappa$ are strictly positive, satisfy the inequality~(\ref{Eq:FinalHypoCausalBound}) and
\begin{equation}
\Gamma_2 =\frac{h}{k_B T}.
\label{Eq:Gamma2}
\end{equation}
\item[(b)] When Eq.~(\ref{Eq:Gamma2}) holds, the characteristic speeds measured by observers comoving with $u^\mu$ are given by $0$, $\pm\sqrt{\mu_0}$, $\pm\sqrt{\mu_1}$, and $\pm\sqrt{\mu_2}$ where
\begin{equation}
\mu_0 = \frac{k_B T}{e}\frac{\eta}{\kappa},\quad
\mu_{1,2} = \frac{1}{2}\left( Tr \pm \sqrt{Tr^2 - 4D} \right),
\label{Eq:Defmu}
\end{equation}
with
\begin{align}
D &= \frac{1}{d}\frac{k_B T}{e},
\label{Eq:DefD}\\
Tr &= \frac{1}{d} + \frac{k_B T}{e}\left[ 1 + 2\left( 1-\frac{1}{d} \right)\frac{\eta}{\kappa} \right].
\label{Eq:DefTr}
\end{align}
\item[(c)] Under the following technical assumptions:
\begin{enumerate}
\item[(i)] $e: (0,\infty)\to (m,\infty)$ is a smooth increasing function satisfying $e(T)\to m > 0$ for $T\to 0$,
\item[(ii)] The heat capacity per particle $c_v = \frac{\partial e}{\partial T}$ is positive and converges to positive values for $T\to 0$ and $T\to \infty$,
\item[(iii)] The speed of sound $v_s := \sqrt{ \frac{k_B T}{h}\frac{c_p}{c_v}}$ satisfies the bounds $\frac{k_B T}{e} \leq v_s^2\leq \frac{1}{d}$,
\item[(iv)] The quantities $\frac{k_B T}{e}\frac{\eta}{\kappa}$ and $\frac{k_B T}{e}\frac{\zeta}{\kappa}$ are bounded and the former converges to a positive value when $T\to \infty$,
\end{enumerate}
all mode solutions proportional to $e^{st + i\vec{k}\cdot\vec{x}}$ with $\vec{k}\neq\vec{0}$ have $\re(s) < 0$, provided that Eq.~(\ref{Eq:Gamma2}) holds and
\begin{equation}
\Gamma_1 = 1 + \frac{c_v}{k_B}\frac{e^2}{k_B T h}\left(\frac{k_B}{c_{v}}-\frac{1}{d}\right)^2\frac{\kappa}{\zeta}\Lambda_0,
\label{Eq:Gamma1}
\end{equation}
with large enough values of the constant $\Lambda_0$. This means that global equilibrium solutions are mode stable.
\item[(d)] The inequality~(\ref{Eq:FinalHypoCausalBound}) and the technical assumptions (i)-(iv) are automatically satisfied for a simple gas whose collisions have a hard sphere or hard disk cross section in $d=3$ and $d=2$ dimensions, respectively.
\end{enumerate}

Furthermore, as shown in App.~\ref{App:Entropy}, the theory described by the system~(\ref{Eq:ndot}--\ref{Eq:qdot}) has an entropy production which is positive up to and including terms which are second order in the gradients.

Note that the choice~(\ref{Eq:Gamma2}) implies $\beta_1 = -k_B T/e$, $\beta_2=0$, and $\beta_3 = -\beta_1/\kappa$, and it eliminates the gradient of the temperature on the right-hand side of Eq.~(\ref{Eq:ConstitutiveRelations_q}), leading to
\begin{align}
\mathcal{Q}_\mu = \kappa \left[\frac{D_\mu n}{n} + \frac{e}{k_B T} a_\mu - \frac{q}{k_B T} E_\mu \right].
\label{Eq:ConstitutiveRelations_qBis}
\end{align}

The main objective of this article is to prove that, under the same hypothesis as in (a) and the assumption that the functions $\mu_0$, $\mu_1$, and $\mu_2$ defined in (b) do not cross each other, the full nonlinear system~(\ref{Eq:ndot}--\ref{Eq:qdot}) leads to a (local in time) well-posed Cauchy problem.

To formulate our main result, we consider a foliation of spacetime $(M,g)$ by spacelike Cauchy hypersurfaces $\Sigma_t$ labeled by $t\in\Real$. We refer to an initial data set as a choice for $(n,T,u^\mu, \epsilon,\mathcal{Q}^\mu)$ on a given Cauchy surface $\Sigma_{t_0}$ which is sufficiently regular\footnote{Here, by ``sufficiently regular" we mean the following. Denote by $X_s(\Sigma_{t_0})$ the space of fields $(n,T,u^\mu,\epsilon,\mathcal{Q}^\mu)$ for which $n$, $T$ and the components of $u^\mu$ lie in the Sobolev space $H^{s+1}(\Sigma_{t_0})$ and $\epsilon$ and the components of $\mathcal{Q}^\mu$ in $H^s(\Sigma_{t_0})$. Then, if  $\Sigma_{t_0}$ is compact, we require the data set to lie in $X_s(\Sigma_{t_0})$ with $s > d/2 + 1$. If $\Sigma_{t_0}$ is not compact, we assume for simplicity that $n$, $T$, $u^\mu$, $\epsilon$ and $\mathcal{Q}^\mu$ are constant outside a compact subset of $\Sigma_{t_0}$ and require instead that the differences with these constant values lie in $X_s(\Sigma_{t_0})$.} and is such that $n,T,\epsilon > 0$, $u^\mu$ is unit timelike and $\mathcal{Q}^\mu$ is orthogonal to $u^\mu$. With these definitions we have:

\begin{theorem}
Assume $e$, $\eta$, $\kappa$, $\zeta$, and $\Gamma_1$ are  smooth ($C^\infty$), strictly positive functions of $T$, fulfilling $0 < c_v = \frac{\partial e}{\partial T} < d k_B$ and either $\Gamma_1 < 1$ or $\Gamma_1 > 1$. Let $\Gamma_2$ be given by~(\ref{Eq:Gamma2}) with $h = e + k_B T$, and suppose the inequality~(\ref{Eq:FinalHypoCausalBound}) is satisfied for all $T > 0$. Furthermore, suppose $\mu_0$, $\mu_1$ and $\mu_2$ are distinct from each other for all $T > 0$.
Let $(n,T,u^\mu, \epsilon,\mathcal{Q}^\mu)$ be an initial data set on $\Sigma_{t_0}$.
Then, there exists $t_1 < t_0$ and $t_2 > t_0$ such that the system~(\ref{Eq:ndot}--\ref{Eq:qdot}) has a unique solution on the submanifold $\bigcup_{t_1 < t < t_2}\Sigma_t$ with the specified initial data on $\Sigma_{t_0}$. Moreover, for each $t\in (t_1,t_2)$, the solution on $\Sigma_t$ depends continuously on the initial data.\footnote{Here, the continuous dependency on the data should be understood with respect to the norm associated with $X_s(\Sigma_t)$ for $s > d/2 + 1$.}
\end{theorem}

The next three sections are devoted to the proof of Theorem~1. The strategy is the following. In the next section we rewrite the evolution equations~(\ref{Eq:ndot}--\ref{Eq:qdot}) in the form of a constrained first-order quasilinear system. Next, in Sec.~\ref{Sec:Hypo} we prove that under the hypotheses of the theorem this system is strongly hyperbolic, which implies by means of standard theorems (see for instance~\cite{Taylor96c}) that it is locally well posed in the Sobolev space $H^s$ with $s > d/2 +1$. Finally, in Sec.~\ref{Sec:ConstraintPropagation} we show that the constraints propagate, which completes the proof of Theorem~1. Because of finite speed of propagation implied by the hyperbolicity property one can localize the problem, that is, one can assume without loss of generality that $\Sigma_{t_0}$ is compact.

\section{First-order quasilinear reformulation}
\label{Sec:FOQ}

In this section we cast the system~(\ref{Eq:ndot}--\ref{Eq:qdot}) as a first-order quasilinear system of equations whose hyperbolicity will be analyzed in the subsequent section. To this purpose, it is necessary to introduce new fields representing, respectively, the spatial gradients of the particle density, the temperature and the velocity field. Associated with these new fields are constraints which must be satisfied to recover the original system.

We can write the constraints as
$C_\mu^{(\mathcal{N})} = 0$, $C_\mu^{(\mathcal{T})} = 0$, and $C_{\mu\nu}^{(B)} = 0$ where
\begin{align}
& C_\mu^{(\mathcal{N})} := \mathcal{N}_\mu - \frac{D_\mu n}{n},
\label{Eq:CN}\\
& C_\mu^{(\mathcal{T})} := \mathcal{T}_\mu - \frac{D_\mu T}{T},
\label{Eq:CT}\\
& C_{\mu\nu}^{(B)} := B_{\mu\nu} - \Delta_\mu{}^\alpha\nabla_\alpha u_\nu,
\label{Eq:CB}
\end{align}
and $\mathcal{N}_\mu$, $\mathcal{T}_\mu$, and $B_{\mu\nu}$ are the new fields representing the aforementioned gradients. Evolution equations for these fields can be obtained by dotting both sides of Eqs.~(\ref{Eq:CN}--\ref{Eq:CB}) and using the commutator identities~(\ref{Eq:Commutator0}) and (\ref{Eq:Dotk}) of Appendix~\ref{App:Commutators}. This yields
\begin{equation}
\dot{C}_\mu^{(\mathcal{N})} = \dot{\mathcal{N}}_\mu - (D_\mu + a_\mu)\left( \frac{\dot{n}}{n} \right)
 + k_\mu{}^\nu\frac{D_\nu n}{n},
\label{Eq:CNdot}
\end{equation}
where $k_{\mu\nu} := \Delta_\mu{}^\alpha\nabla_\alpha u_\nu$, an analogous equation for $\dot{C}_\mu^{(\mathcal{T})}$, and
\begin{equation}
\dot{C}_{\mu\nu}^{(B)} = \dot{B}_{\mu\nu} - (D_\mu + a_\mu)a_\nu + k_\mu{}^\alpha k_{\alpha\nu} + R_{\mu\alpha\nu\beta} u^\alpha u^\beta,
\label{Eq:CBdot}
\end{equation}
where here $R_{\mu\alpha\nu\beta}$ denotes the Riemann curvature tensor. Imposing the constraints and substituting $\dot{n}/n$, $\dot{T}/T$ and $a_\mu$ with the right-hand sides of Eqs.~(\ref{Eq:ndot}), (\ref{Eq:TdotBis}) and (\ref{Eq:udotBis}), respectively, one obtains the following evolution equations for the new fields:
\begin{align}
\dot{\mathcal{N}}_\mu &= -(D_\mu + a_\mu)\theta - B_\mu{}^\nu\mathcal{N}_\nu,
\label{Eq:dotmathcalN}\\
\dot{\mathcal{T}}_\mu &= \alpha_4 (D_\mu + a_\mu)\theta + \alpha_5 (D_\mu + a_\mu)\epsilon
 + \mathcal{T}_\mu\Pi^{(\alpha)}
\nonumber\\
 &- B_\mu{}^\nu\mathcal{T}_\nu,
\label{Eq:dotmathcalT}\\
\dot{B}_{\mu\nu} &= 
\beta_1 D_\mu\mathcal{N}_\nu 
+ \beta_2 D_\mu\mathcal{T}_\nu 
+ \beta_3 D_\mu\mathcal{Q}_\nu 
\nonumber\\
 &- B_\mu{}^\alpha B_{\alpha\nu} 
  - R_{\mu\alpha\nu\beta} u^\alpha u^\beta + a_\mu a_\nu\nonumber\\
 & -\frac{q\beta_{1}}{k_B T}D_{\mu}E_{\nu}+\mathcal{T_{\mu}}\left(\Pi_{\nu}^{\left(\beta\right)}+\Pi_{\nu}^{\left(E\right)}\right),
\label{Eq:dotB}
\end{align}
where we have defined 
\begin{eqnarray}
\Pi^{(\alpha)} &:=& T\left[ \frac{\partial\alpha_4}{\partial T}\theta + \frac{\partial\alpha_5}{\partial T}\epsilon \right],
\label{Eq:Pialpha}\\
\Pi^{(\beta)}_\nu &:=& T\left[ \frac{\partial\beta_1}{\partial T}\mathcal{N}_\nu
+ \frac{\partial\beta_2}{\partial T}\mathcal{T}_\nu
+ \frac{\partial\beta_3}{\partial T} \mathcal{Q}_\nu \right],
\label{Eq:Pibeta}\\
\Pi_{\nu}^{\left(E\right)}&:=&\left[\beta_{1}-T\frac{\partial\beta_{1}}{\partial T}\right]\frac{q E_{\nu}}{k_{B}T},
\label{Eq:PiE}
\end{eqnarray}
and used the fact that the coefficients $\alpha_i$ and $\beta_i$ depend on the temperature only. The terms involving the acceleration $a_\mu$ in Eqs.~(\ref{Eq:dotmathcalN}--\ref{Eq:dotB}) should be substituted by the right-hand side of Eq.~(\ref{Eq:udotBis}). Together with Eqs.~(\ref{Eq:ndot}--\ref{Eq:qdot}), where $D_\mu n/n$, $D_\mu T/T$, $\sigma_{\mu\nu}$, $\omega_{\mu\nu}$, and $\theta$ are replaced by $\mathcal{N}_\mu$, $\mathcal{T}_\mu$, $B_{\langle \mu\nu \rangle}$, $B_{[\mu\nu]}$ and $\Delta^{\mu\nu} B_{\mu\nu}$, respectively, the equations~(\ref{Eq:dotmathcalN}-\ref{Eq:dotB}) yield a first-order quasilinear evolution system for the fields $U := (n,T,u_\mu,\mathcal{N}_\mu,\mathcal{T}_\mu,\mathcal{Q}_\mu,B_{\mu\nu},\epsilon)$ which is subject to the constraints $C_\mu^{(\mathcal{N})} = C_\mu^{(\mathcal{T})} = 0$ and $C_{\mu\nu}^{(B)} = 0$. However, as we will understand shortly, the resulting system is not strongly hyperbolic off the constraint hypersurface, that is, when considering solutions which (slightly) violate the constraints. At this point the reader might wonder why it is necessary to worry about solutions that do not satisfy the constraints, since they are unphysical. The reason is that, from a practical point of view, it is important to understand the behavior of small violations of the constraints, in particular when considering solving the first-order system numerically. Unless the numerical discretization guarantees that the constraints are {\it exactly} satisfied, numerical errors push the solution curve off the constraint hypersurface, and thus one needs to first understand the dynamical behavior of the first-order system without taking into account the constraints.

For this reason, it is important to obtain a first-order system which is strongly hyperbolic not just on the constraint hypersurface, but also off it (or at least in a vicinity of it). As we will see, this can be achieved by adding terms which are linear in the constraint fields $C_\mu^{(\mathcal{N})}$, $C_\mu^{(\mathcal{T})}$ and $C_{\mu\nu}^{(B)}$ and their first-order antisymmetric derivatives to the right-hand side of the first-order system. Indeed, this addition does not alter the solution flow on the constraint hypersurface, although it modifies the dynamics {\it off} this surface. The trick is to exploit this freedom in such a way that the new system is strongly hyperbolic on and off the constraint hypersurface.

Using the identities~(\ref{Eq:CommutatorBis0}) and (\ref{Eq:Curlk}) and the definitions~(\ref{Eq:CN}-\ref{Eq:CB}) one finds
\begin{align}
D_{[\mu} C^{(\mathcal{N})}_{\nu]} &= D_{[\mu} \mathcal{N}_{\nu]} - k_{[\mu\nu]}\frac{\dot{n}}{n},\\
D_{[\mu} C^{(\mathcal{T})}_{\nu]} &= D_{[\mu} \mathcal{T}_{\nu]} - k_{[\mu\nu]}\frac{\dot{T}}{T},\\
D_{[\mu} C^{(B)}_{\nu]\alpha} &= D_{[\mu} B_{\nu]\alpha} - k_{[\mu\nu]} a_\alpha - \frac{1}{2}\Delta_\mu{}^{\mu'}\Delta_\nu{}^{\nu'} R_{\mu'\nu'\alpha\beta} u^\beta.
\end{align}
The derivatives of $u^\mu$ that appear in $k_{[\mu\nu]}$ can be replaced by $B_{[\mu\nu]}$ using the constraint field $C^{(B)}_{\mu\nu}$. This gives rise to the additional constraint fields
\begin{align}
Z^{(\mathcal{N})}_{\mu\nu} &:= D_{[\mu} C^{(\mathcal{N})}_{\nu]} - C_{[\mu\nu]}^{(B)}\frac{\dot{n}}{n}
= D_{[\mu} \mathcal{N}_{\nu]} - B_{[\mu\nu]}\frac{\dot{n}}{n},
\label{Eq:ZmathcalN}\\
Z^{(\mathcal{T})}_{\mu\nu} &:= D_{[\mu} C^{(\mathcal{T})}_{\nu]} - C_{[\mu\nu]}^{(B)}\frac{\dot{T}}{T}
= D_{[\mu} \mathcal{T}_{\nu]} - B_{[\mu\nu]}\frac{\dot{T}}{T},
\label{Eq:ZmathcalT}\\
Z^{(B)}_{\mu\nu\alpha} &:= D_{[\mu} C^{(B)}_{\nu]\alpha}
 - C^{(B)}_{[\mu\nu]} a_\alpha
\nonumber\\
 &= D_{[\mu} B_{\nu]\alpha} - B_{[\mu\nu]} a_\alpha - \frac{1}{2}\Delta_\mu{}^{\mu'}\Delta_\nu{}^{\nu'} R_{\mu'\nu'\alpha\beta} u^\beta,
\label{Eq:ZBN}
\end{align}
where here, one should substitute $\dot{n}/n$ and $\dot{T}/T$ with the right-hand sides of Eqs.~(\ref{Eq:ndot}) and (\ref{Eq:TdotBis}), respectively.

Contracting Eq.~(\ref{Eq:ZBN}) over the indices $\mu$ and $\alpha$ one obtains $Z_\nu := 2\Delta^{\mu\alpha} Z^{(B)}_{\mu\nu\alpha}$ with
\begin{align}
Z_\mu &= D^\alpha C^{(B)}_{\mu\alpha} - D_\mu C^{(\theta)} -2 C^{(B)}_{[\alpha\mu]} a^\alpha
\nonumber\\
 &= D^\alpha B_{\mu\alpha} - D_\mu\theta
 - 2\omega_{\alpha\mu} a^\alpha - \Delta_\mu{}^{\mu'} R_{\mu'\alpha} u^\alpha,
\label{Eq:Z}
\end{align}
where we have abbreviated $C^{(\theta)} := \Delta^{\alpha\beta} C^{(B)}_{\alpha\beta}$. Note that from now on, $\theta = \Delta^{\alpha\beta}B_{\alpha\beta}$, $\sigma_{\mu\nu} = B_{\langle \mu\nu \rangle}$ and $\omega_{\mu\nu} = B_{[\mu\nu]}$ should be considered as independent variables which only agree with the divergence, shear and vorticity associated with $u^\mu$ on the constraint hypersurface.

With these definitions, the first-order system for $U$ whose hyperbolicity will be analyzed in the next section can be written as follows:
\begin{align}
\dot{n} &= -\theta n,
\label{Eq:FOndot}\\
\dot{T} &= (\alpha_4\theta + \alpha_5\epsilon)T,
\label{Eq:FOTdot}\\
\dot{u}_\mu &= \beta_1\mathcal{N}_\mu + \beta_2\mathcal{T}_\mu + \beta_3 \mathcal{Q}_\mu -\frac{q \beta_{1}}{k_{B}T}E_{\mu},
\label{Eq:FOudot}\\
\dot{\mathcal{N}}_\mu &= -(D_\mu + a_\mu)\theta - B_\mu{}^\nu\mathcal{N}_\nu
 + \delta_1 Z_\mu,
\label{Eq:FOmathcalNdot}\\
\dot{\mathcal{T}}_\mu &= \alpha_4 (D_\mu + a_\mu)\theta + \alpha_5 (D_\mu + a_\mu)\epsilon 
 + \mathcal{T}_\mu\Pi^{(\alpha)}
\nonumber\\
 &- B_\mu{}^\nu\mathcal{T}_\nu
 + \delta_2 Z_\mu,
\label{Eq:FOmathcalTdot}\\
\dot{\mathcal{Q}}_\mu &= 2\eta D^\nu\sigma_{\mu\nu} - \frac{1}{d} D_\mu\epsilon + (f_\mathcal{Q})_\mu
 + \delta_3 Z_\mu,
\label{Eq:FOqdot}\\
\dot{B}_{\mu\nu} &= 
\beta_1 D_\mu\mathcal{N}_\nu 
+ \beta_2 D_\mu\mathcal{T}_\nu 
+ \beta_3 D_\mu \mathcal{Q}_\nu-\frac{q\beta_{1}}{k_B T}D_{\mu}E_{\nu} \nonumber\\
 &- B_\mu{}^\alpha B_{\alpha\nu} 
  - R_{\mu\alpha\nu\beta} u^\alpha u^\beta  + a_\mu a_\nu 
\nonumber\\
&+\mathcal{T_{\mu}}\left(\Pi_{\nu}^{\left(\beta\right)}+\Pi_{\nu}^{\left(E\right)}\right)- \beta_1 Z^{(\mathcal{N})}_{\mu\nu} 
 - \beta_2 Z^{(\mathcal{T})}_{\mu\nu},
\label{Eq:FOBdot}\\
\dot{\epsilon} &= -D^\mu \mathcal{Q}_\mu + f_{\epsilon},
\label{Eq:FOEdot}
\end{align}
where $(f_\mathcal{Q})_\mu$ and $f_{\epsilon}$ depend pointwise (i.e. algebraically) on  $U = (n,T,u_\mu,\mathcal{N}_\mu,\mathcal{T}_\mu,\mathcal{Q}_\mu,B_{\mu\nu},\epsilon)$. In these expressions, it is understood that all terms $a_\mu$ are to be substituted with the right-hand side of Eq.~(\ref{Eq:FOudot}). Furthermore, the functions $\delta_1$, $\delta_2$ and $\delta_3$ (which should depend only on $T$) will be adjusted in the next section to determine how the constraint fields are added to the right-hand sides of the evolution equations for $\mathcal{N}_\mu$, $\mathcal{T}_\mu$ and $\mathcal{Q}_\mu$. In turn, the constraints $Z_{\mu\nu}^{(\mathcal{N})}$ and $Z_{\mu\nu}^{(\mathcal{T})}$ are added to the evolution equation for $B_{\mu\nu}$ with the coefficients $-\beta_1$ and $-\beta_2$, respectively. The reason for this becomes clear when rewriting this equation in terms of its trace, symmetric traceless and antisymmetric parts, which yields (cf. Eqs.~(\ref{Eq:Raychaudhuri}--\ref{Eq:VorticityProp}) in Appendix~\ref{App:Commutators})
\begin{align}
\dot{\theta}&=\beta_{1}D^{\mu}\mathcal{N}_{\mu}+\beta_{2}D^{\mu}\mathcal{T}_{\mu}+\beta_{3}D^{\mu}\mathcal{Q}_{\mu}-\frac{q\beta_{1}}{k_B T}D^{\mu}E_{\mu}\nonumber\\ 
&-\sigma^{2}+\omega^{2}-\frac{\theta^{2}}{d}-R_{\mu\nu}u^{\mu}u^{\nu}+a^{\mu}a_{\mu}\nonumber\\
&+\mathcal{T}^{\nu}\left(\Pi_{\nu}^{(\beta)}+\Pi_{\nu}^{\left(E\right)}\right),
\label{Eq:FOthetadot}\\
\dot{\sigma}_{\mu\nu}&=\beta_{1}D_{\langle\mu}\mathcal{N}_{\nu\rangle}+\beta_{2}D_{\langle\mu}\mathcal{T}_{\nu\rangle}+\beta_{3}D_{\langle\mu}\mathcal{Q}_{\nu\rangle}-\frac{q\beta_{1}}{k_B T}D_{\langle\mu}E_{\nu\rangle}\nonumber\\ 
&-\sigma_{\langle\mu}{}^{\alpha}\sigma_{\nu\rangle\alpha}+\omega_{\langle\mu}{}^{\alpha}\omega_{\nu\rangle\alpha}-\frac{2\theta}{d}\sigma_{\mu\nu}+a_{\langle\mu}a_{\nu\rangle}\nonumber\\
&-\left(R_{\mu\alpha\nu\beta}-\frac{1}{d}\Delta_{\mu\nu}R_{\alpha\beta}\right)u^{\alpha}u^{\beta}+\mathcal{T_{\langle\mu}}\left(\Pi_{\nu\rangle}^{\left(\beta\right)}+\Pi_{\nu\rangle}^{\left(E\right)}\right),
\label{Eq:FOsigmadot}\\
\dot{\omega}_{\mu\nu}&=\beta_{3}D_{[\mu}\mathcal{Q}_{\nu]}-\frac{q\beta_{1}}{k_{B}T}D_{[\mu}E_{\nu]}+2\sigma_{[\mu}{}^{\alpha}\omega_{\nu]\alpha}\nonumber\\
&+\omega_{\mu\nu}\left[\beta_{2}(\alpha_{4}\theta+\alpha_{4}\epsilon)-\theta\left(\beta_{1}+\frac{2}{d}\right)\right]\nonumber\\
&+\mathcal{T}_{[\mu}\left(\Pi_{\nu]}^{\left(\beta\right)}+\Pi_{\nu]}^{\left(E\right)}\right),
\label{Eq:FOomegadot}
\end{align}
where we have abbreviated $\sigma^2 := \sigma^{\mu\nu}\sigma_{\mu\nu}$ and $\omega^2 := \omega^{\mu\nu}\omega_{\mu\nu}$. On the constraint hypersurface, the first equation reduces to the Raychaudhuri equation when the acceleration is zero, and it determines the time evolution of the expansion of the fluid, whereas the second equation determines the evolution of the shear. The last equation describes the evolution of the vorticity of the fluid, and now the reason for having subtracted the last two terms in Eq.~(\ref{Eq:FOBdot}) become clear: they eliminate the antisymmetric derivatives of $\mathcal{N}_\mu$ and $\mathcal{T}_\mu$ that would otherwise appear in the evolution equation for $\omega_{\mu\nu}$. Consequently, only derivatives of the heat flux appear in this equation, which simplifies the principal part of the equations.

\section{Strong hyperbolicity}
\label{Sec:Hypo}

The first-order evolution system~(\ref{Eq:FOndot}-\ref{Eq:FOEdot}) or, equivalently, Eqs. (\ref{Eq:FOndot}-\ref{Eq:FOqdot},\ref{Eq:FOthetadot}-\ref{Eq:FOomegadot},\ref{Eq:FOEdot}) has the form
\begin{equation}
\dot{U} = \mathcal{A}^\mu(U) D_\mu U + \mathcal{F}(U),
\label{Eq:FOS}
\end{equation}
where the matrices $\mathcal{A}^\mu(U)$, $\mu=0,1,2,\ldots,d$, determine the coefficients in front of the spatial derivatives of the state vector $U$ and $\mathcal{F}(U)$ is a nonlinear algebraic function of $U$. Before we proceed, it is important to recall that $\dot{U}$ refers to the directional derivative of $U$ along the velocity field $u^\mu$ and likewise, $D_\mu$ denotes the derivatives in the directions orthogonal to $u^\mu$. Hence, when rewriting Eq.~(\ref{Eq:FOS}) as a PDE system with respect to some local coordinates $(t,x^1,\ldots,x^d)$ adapted to a foliation by $t=const$ spacelike hypersurfaces, time derivatives will appear on the right-hand side of the equations.\footnote{Otherwise, the operators $D_\mu$ would be tangent to the $t=const$ hypersurfaces which means that the vector field $u^\mu$ is hypersurface orthogonal. However, this would imply that the vorticity vanishes identically, $\omega_{\mu\nu} = 0$, which in general would be inconsistent with Eq.~(\ref{Eq:FOomegadot}) which shows that the heat flux and the electric field can generate vorticity of the particle flow.} Therefore, to compute the time derivatives of $U$ (as required, for example, when implementing a numerical scheme based on the method of lines), one needs to invert a matrix. The details and validity of this inversion are analyzed in Appendix~\ref{App:StronglyHyperbolicPDE}. Fortunately, one can use the methods developed in Refs.~\cite{rG96,oR04} (see also \cite{oSeBjP19}) in order to determine the hyperbolicity of the system~(\ref{Eq:FOS}) without rewriting it explicitly as a PDE system, and this greatly simplifies the analysis. For this, one considers the {\it principal symbol}, defined as
\begin{equation}
\mathcal{A}(k,U) := \mathcal{A}^\mu(U) k_\mu,
\label{Eq:PrincipalSymbol}
\end{equation}
where $k_\mu$ is a given covector perpendicular to $u^\mu$. The first-order system~(\ref{Eq:FOS}) is called (cf. Definition~2 in~\cite{oSeBjP19})
\begin{itemize}
\item {\it strongly hyperbolic} if there exists a symmetric, positive definite matrix $H(k,U)$ (called the symmetrizer), depending smoothly on $k$ and $U$, such that
\begin{equation}
H(k,U)\mathcal{A}(k,U)
\label{Eq:SH}
\end{equation}
is symmetric for all $k$ and $U$,
\item {\it symmetric hyperbolic} if it is strongly hyperbolic and the symmetrizer can be chosen independent of $k$.
\end{itemize}
In App.~\ref{App:StronglyHyperbolicPDE} we show that the above definitions, together with an additional condition on the eigenvalues of the principal symbol, lead to hyperbolic PDEs for which standard theorems guarantee that the Cauchy problem is locally well posed~\cite{Kreiss89,Taylor96c}. This means that, given initial data $U(t_0)$ at time $t_0$, there exists a unique solution $U(t)$, defined on some time interval $(t_1,t_2)$ containing $t_0$, which depends continuously on the initial data with respect to the Sobolev space $H^s$ with $s > d/2 + 1$.

Note that the condition~(\ref{Eq:SH}) implies that $\mathcal{A}(k,U)$ is symmetric with respect to the scalar product defined by $H(k,U)$ and hence it is diagonalizable and has only real eigenvalues. Conversely, if $\mathcal{A}(k,U)$ is diagonalizable and has a real spectrum, and if $S(k,U)$ denotes the matrix whose columns are the eigenvectors of $\mathcal{A}(k,U)$, then it is easy to check that
\begin{equation}
H(k,U) := [S(k,U)^{-1}]^T S(k,U)^{-1}
\end{equation}
is a symmetrizer, and it only remains to verify that it depends smoothly on $k$ and $U$ for the system to be strongly hyperbolic.

For the following we explicitly construct a symmetrizer along the lines we have just sketched. In a first step, we note that the principal symbol exhibits a particular block structure that simplifies the analysis. Next, we show that the symbol can be decoupled into a scalar, vector and tensor block which are independent of $k$. Finally, the symmetrizer is constructed by diagonalizing each of these blocks. Without loss of generality we assume from now on that $k_\mu$ is normalized such that $k^\mu k_\mu = 1$.

\subsection{Block structure}

The analysis of the system~(\ref{Eq:FOndot}--\ref{Eq:FOEdot}) is simplified by observing that it possesses the following block structure: decompose $U = (U_0,U_1,U_2)$ according to
\begin{equation}
U_0 := \begin{pmatrix}
        n \\ T \\ u_\mu
    \end{pmatrix},\quad
U_1 := \begin{pmatrix}
        \mathcal{N}_\mu \\
        \mathcal{T}_\mu \\
        \mathcal{Q}_\mu
    \end{pmatrix}, \quad
U_2 := \begin{pmatrix}
        \theta \\
        \sigma_{\mu \nu} \\
        \omega_{\mu \nu} \\
        \epsilon
    \end{pmatrix}.
\end{equation}
Then, with respect to this decomposition the principal symbol has the form
\begin{equation}
\mathcal{A}(k,U) = \begin{pmatrix}
    0 & 0 & 0 \\
    0 & 0 & Q(k,U_0) \\
    0 & R(k,U_0) & 0
    \end{pmatrix},
\end{equation}
where the symbols $Q(k,U_0)$ and $R(k,U_0)$ are given by
\begin{align}
&Q(k,U_0) U_2 = \begin{pmatrix}
    -k_\mu\theta \\
    \alpha_4 k_\mu\theta + \alpha_5 k_\mu\epsilon 
    \\
    2\eta k^\nu\sigma_{\mu\nu} - \frac{1}{d} k_\mu\epsilon
\end{pmatrix}
\nonumber\\
 &+ \left[ k^\nu(\sigma_{\mu\nu} + \omega_{\mu\nu}) - \left(1-\frac{1}{d} \right)k_\mu\theta\right]
 \begin{pmatrix} 
 \delta_1 \\ \delta_2 \\ \delta_3
 \end{pmatrix},
\end{align}
and
\begin{equation}
R(k,U_0) U_1 = \begin{pmatrix}
\beta_1 k^\mu\mathcal{N}_\mu 
+ \beta_2 k^\mu\mathcal{T}_\mu
+ \beta_3 k^\mu \mathcal{Q}_\mu \\
 \beta_1 k_{\langle \mu}\mathcal{N}_{\nu \rangle} + \beta_2 k_{\langle \mu}\mathcal{T}_{\nu \rangle} + \beta_3 k_{\langle \mu}\mathcal{Q}_{\nu \rangle} \\
\beta_3 k_{[\mu} \mathcal{Q}_{\nu]} \\
 -k^\mu \mathcal{Q}_\mu
\end{pmatrix},
\end{equation}
and depend only on the state variables $U_0 = (n,T,u^\mu)$ and not the full vector $U$.\footnote{The dependency on $U_0$ arises because of the fact that the coefficients $\alpha_i$, $\beta_i$ and $\eta$ depend on the temperature $T$ and $k_\mu$ is orthogonal to $u^\mu$.}

\subsection{Decoupling of the scalar, vector and tensor modes}

The principal symbol can be further simplified by decomposing the fields in terms of components parallel and orthogonal to $k_\mu$. More precisely, let $e^0_\mu$, $e^1_\mu,\ldots,e^d_\mu$ be an orthonormal frame of covector fields such that $e^0_\mu = u_\mu$ and $e^1_\mu = k_\mu$. Then, we may decompose
\begin{equation}
\mathcal{N}_\mu = \mathcal{N}^{\parallel} k_\mu + \mathcal{N}^{\perp}_A e^A_\mu,\qquad
A = 2,\dots,d,
\end{equation}
and similarly for $\mathcal{T}_\mu$ and $\mathcal{Q}_\mu$. Likewise, we decompose
\begin{align}
\sigma_{\mu\nu} &= \left( k_\mu k_\nu - \frac{1}{d}\Delta_{\mu\nu} \right)\sigma_{\parallel} 
+ \sigma_{A}^\perp k_{(\mu} e^A_{\nu)}
 + \sigma^{\perp\perp}_{AB} e^A_\mu e^B_{\nu},
\\
\omega_{\mu\nu} &= 
 \omega_A^\perp k_{[\mu} e^A_{\nu]} + \omega^{\perp\perp}_{AB} e^A_\mu e^B_\nu,\qquad
A,B = 2,\dots,d,
\end{align}
where $\sigma_{AB}$ and $\omega_{AB}$ are symmetric and antisymmetric in $AB$, respectively, and 
$\delta^{AB}\sigma^{\perp\perp}_{AB} = 0$. This decomposition defines a linear and invertible map
\begin{equation}
T(k): (U_0,U_1,U_2)\mapsto (U_0,U_2^{\perp\perp},U_1^\parallel,U_2^\parallel,U_1^\perp,U_2^\perp), 
\end{equation}
where
\begin{align}
& U_1^\parallel = \begin{pmatrix}
        \mathcal{N}^\parallel \\
        \mathcal{T}^\parallel \\
        \mathcal{Q}^\parallel
    \end{pmatrix}, \quad
U_1^\perp = \begin{pmatrix}
        \mathcal{N}^\perp \\
        \mathcal{T}^\perp \\
        \mathcal{Q}^\perp
    \end{pmatrix},
\\
& U_2^\parallel = \begin{pmatrix}
        \theta \\
        \sigma^\parallel \\
        \epsilon
    \end{pmatrix},\quad
U_2^\perp = \begin{pmatrix}
     \sigma^\perp \\
     \omega^\perp
    \end{pmatrix},\quad
U_2^{\perp\perp} = \begin{pmatrix}
     \sigma^{\perp\perp} \\
     \omega^{\perp\perp}
    \end{pmatrix}.
\end{align}
The transformed symbol
$\tilde{\mathcal{A}}(U) := T(k)\mathcal{A}(k,U) T(k)^{-1}$ has the following block form:
\begin{equation}
\tilde{\mathcal{A}}(U) = \begin{pmatrix}
  0 & 0 & 0 & 0 & 0 & 0\\
  0 & 0 & 0 & 0 & 0 & 0\\
  0 & 0 & 0 & Q^\parallel(U_0) & 0 & 0\\
  0 & 0 & R^\parallel(U_0) & 0 & 0 & 0\\
  0 & 0 & 0 & 0 & 0 & Q^\perp(U_0) \\
  0 & 0 & 0 & 0 & R^\perp(U_0) & 0
\end{pmatrix},
\label{Eq:SymbolScalarVecTensorForm}
\end{equation}
and hence the {\it scalar block} (consisting of the fields $U_1^\parallel$ and $U_2^\parallel$), the {\it vector block} (consisting of $U_1^\perp$ and $U_2^\perp$) and the {\it tensor block} (consisting of $U_2^{\perp\perp}$) decouple from each other.
The tensor block is trivial, whereas the matrices in the scalar block are
\begin{align}
Q^\parallel(U_0) &= \begin{pmatrix}
    -1 - \delta_1\left(1 - \frac{1}{d}\right)          &   \delta_1\left(1 - \frac{1}{d}\right)       &   0 \\
    \alpha_4 -\delta_2\left(1 - \frac{1}{d}\right)    &   \delta_2\left(1 - \frac{1}{d}\right)      &   \alpha_5 \\
     -\delta_3\left(1- \frac{1}{d}\right)           & \left(1- \frac{1}{d}\right)(2\eta+\delta_3) &   -\frac{1}{d}
    \end{pmatrix},
\label{Eq:Qparallel}\\
R^\parallel(U_0) &= \begin{pmatrix}
    \beta_1 & \beta_2 &   \beta_3 \\
    \beta_1 & \beta_2 &   \beta_3 \\
        0   & 0  &   -1
    \end{pmatrix},
\label{Eq:Rparallel}
\end{align}
and in the vector block
\begin{align}
Q^\perp(U_0) &= \frac{1}{2}
\begin{pmatrix}
    \delta_1 & -\delta_1 \\
    \delta_2 & -\delta_2 \\
    \delta_3 + 2\eta & -\delta_3 
\end{pmatrix},
\label{Eq:Qperp}\\
R^\perp(U_0) &= 
\begin{pmatrix}
    \beta_1 & \beta_2 & \beta_3 \\
    0   & 0 & \beta_3
\end{pmatrix}.
\label{Eq:Rperp}
\end{align}
Note that the transformed symbols $\tilde{\mathcal{A}}(U)$, $Q^\parallel(U_0)$, $R^\parallel(U_0)$, $Q^\perp(U_0)$, $R^\perp(U_0)$ are independent of $k_\mu$.

\subsection{Diagonalizing the vector block}

We start with the vector block consisting of the matrix
\begin{equation}
\tilde{\mathcal{A}}^\perp(U) := \begin{pmatrix}
  0 & Q^\perp(U_0) \\
  R^\perp(U_0) & 0
\end{pmatrix},
\label{Eq:Atildeperp}
\end{equation}
with $Q^\perp(U_0)$ and $R^\perp(U_0)$ given in Eqs.~(\ref{Eq:Qperp},\ref{Eq:Rperp}). For the following analysis we use:

\begin{lemma}
\label{Lem:RQ}
Let $n,m$ be natural numbers and let $Q$ and $R$ be $n\times m$ and $m\times n$ matrices, respectively. Let $\mu$ be a nonzero eigenvalue of $M := R Q$. Then, $\pm\sqrt{\mu}$ are eigenvalues of
\begin{equation}
A := \begin{pmatrix}
 0 & Q \\
 R & 0
\end{pmatrix}.
\end{equation}
Furthermore, if $M$ is diagonalizable and invertible, then $A$ is diagonalizable and its spectrum has the form
\begin{equation}
\sigma(A) = \{ 0,\pm\sqrt{\mu_1},\ldots,\pm\sqrt{\mu_s} \},
\end{equation}
where $\mu_1,\ldots,\mu_s$ are the eigenvalues of $M$. If $n=m$, then the zero eigenvalue is omitted.
\end{lemma}

\proof The proof is similar to the one of Lemma~1 in Ref. ~\cite{oSeBjP19}, so we only sketch it. Suppose $v_1$ and $v_2$ are linearly independent eigenvectors of $M$ with nonzero eigenvalues $\mu_1$ and $\mu_2$. Then it is not difficult to verify that the $4$ vectors
\begin{equation}
\left( \begin{array}{c} \pm\frac{1}{\sqrt{\mu_i}} Q v_i \\ v _i \end{array} \right),\qquad i =1,2,
\end{equation}
are linearly independent eigenvectors of $A$ with eigenvalues $\pm\sqrt{\mu_i}$. The eigenvectors of $A$ which are not of this form (if present) have eigenvalue $0$ and are constructed from the elements of the kernels of $Q$ and $R$.
\qed

To apply the lemma to the symbol defined in Eq.~(\ref{Eq:Atildeperp}) we compute
\begin{equation}
R^\perp(U_0) Q^\perp(U_0) = 
\frac{1}{2}\begin{pmatrix}
 \vec{\beta}\cdot\vec{\delta} + 2\beta_3\eta & -\vec{\beta}\cdot\vec{\delta} \\
 \beta_3(\delta_3 + 2\eta) &  -\beta_3\delta_3 
\end{pmatrix},
\end{equation}
where we have introduced the shorthand notation $\vec{\beta}\cdot\vec{\delta} :=\beta_1\delta_1 + \beta_2\delta_2 + \beta_3\delta_3$. The eigenvalues of this matrix are
\begin{equation}
\mu_1^\perp:=\frac{1}{2}\left( \beta_1\delta_1 + \beta_2\delta_2 \right),\qquad
\mu_2^\perp := \beta_3\eta,
\end{equation}
and according to the lemma it follows that $\tilde{\mathcal{A}}^\perp(U)$ is diagonalizable with only real eigenvalues if $\mu_1^\perp$ and $\mu_2^\perp$ are strictly positive and $R^\perp(U_0) Q^\perp(U_0)$ is diagonalizable. A particularly attractive possibility to achieve this is choosing $\delta_1$, $\delta_2$, and $\delta_3$, such that
\begin{equation}
\vec{\beta}\cdot\vec{\delta} = 0,\qquad
\delta_3 = -2\eta,
\label{Eq:deltaChoice1}
\end{equation}
which yields $R^\perp(U_0) Q^\perp(U_0) = \eta\beta_3 I$. Hence, if $\beta_3\eta > 0$ and Eq.~(\ref{Eq:deltaChoice1}) holds, the vector block is diagonalizable with the purely real eigenvalues:
\begin{equation}
0,\quad \pm\sqrt{\beta_3\eta}.
\end{equation}
Furthermore, the system is causal if $0 < \beta_3\eta\leq 1$. This is the same condition that was found in Eq.~(20) of the companion letter~\cite{aGjSoS2024a} when analyzing the vector block of the second-order linearized system, which motivated the choice for $\Gamma_2$ in Eq. (\ref{Eq:Gamma2})

For simplicity, we shall stick to the choice~(\ref{Eq:deltaChoice1}) for what follows.

\subsection{Diagonalizing the scalar block}

Next, we turn our attention to the scalar block consisting of the matrix
\begin{equation}
\tilde{\mathcal{A}}^\parallel(U) := \begin{pmatrix}
  0 & Q^\parallel(U_0) \\
  R^\parallel(U_0) & 0
\end{pmatrix},
\end{equation}
with $Q^\parallel(U_0)$ and $R^\parallel(U_0)$ given in Eqs.~(\ref{Eq:Qparallel},\ref{Eq:Rparallel}). To analyze under which conditions $\tilde{\mathcal{A}}^\parallel(U)$ is diagonalizable with only real eigenvalues we use again Lemma~\ref{Lem:RQ}, where in this case the matrix $M^\parallel(U_0) := R^\parallel(U_0) Q^\parallel(U_0)$ is
\begin{equation}
M^\parallel(U_0)
= \begin{pmatrix}
 \alpha_4\beta_2-\beta_1 & A_0\beta_3 & \alpha_5\beta_2-\frac{\beta_3}{d} \\
 \alpha_4\beta_2-\beta_1 & A_0\beta_3 & \alpha_5\beta_2-\frac{\beta_3}{d} \\
 -A_0 & 0 & \frac{1}{d}
\end{pmatrix},
\end{equation}
and we have abbreviated $A_0 := 2(1 - 1/d)\eta$ and imposed the condition~(\ref{Eq:deltaChoice1}). Next, we claim that, apart from the zero eigenvalue, this matrix has precisely the same eigenvalues as the matrix $R Q$ obtained from Eq.~(26) in~\cite{aGjSoS2024a}, which is
\begin{equation}
\tilde{M}^\parallel := \begin{pmatrix}
\alpha_4\beta_2  -\beta_1 + A_0\beta_3 
& \alpha_5\beta_2 - \frac{\beta_3}{d} \\
 - A_0 & \frac{1}{d} 
\end{pmatrix}.
\end{equation}
To prove this statement, we compute the characteristic polynomial $p_{M^\parallel}(\lambda)$ corresponding to $M^\parallel(U_0)$, which gives
\begin{align}
p_{M^\parallel}(\lambda) & = \begin{vmatrix}
\alpha_4\beta_2-\beta_1 - \lambda & A_0\beta_3 & \alpha_5\beta_2-\frac{\beta_3}{d} \\
    \lambda &  - \lambda & 0 \\
    -A_0 & 0 & \frac{1}{d} - \lambda
\end{vmatrix}
\nonumber\\
 &= \begin{vmatrix}
\alpha_4\beta_2-\beta_1 + A_0\beta_3 - \lambda & A_0\beta_3 & \alpha_5\beta_2-\frac{\beta_3}{d} \\
    0 &  - \lambda & 0 \\
    -A_0 & 0 & \frac{1}{d} - \lambda
\end{vmatrix} 
\nonumber\\
 &= -\lambda p_{\tilde{M}^\parallel}(\lambda),
\end{align}
where in the first step we have subtracted the first row from the second one, and in the second step we have added the second column to the first one. This proves the claim.

In the companion letter~\cite{aGjSoS2024a} we proved that the eigenvalues $\mu_1$ and $\mu_2$ of the matrix $\tilde{M}^\parallel$ satisfy $0 < \mu_1 < \mu_2 \leq 1$ if Eq.~(\ref{Eq:Gamma2}) and the inequality~(\ref{Eq:FinalHypoCausalBound}) hold. Explicitly, they are given by
\begin{equation}
\mu_{1,2} = \frac{1}{2}\left( Tr \pm \sqrt{Tr^2 - 4D} \right),
\end{equation}
where
\begin{equation}
D = \frac{1}{d}\frac{k_B T}{e},\qquad
Tr = \frac{1}{d} + \frac{k_B T}{e}\left( 1 + \frac{A_0}{\kappa} \right).
\end{equation}
According to the proof of Lemma~\ref{Lem:RQ}, two distinct nonzero eigenvalues of $\tilde{M}^\parallel$ give rise to four linearly independent eigenvectors of the matrix $\tilde{\mathcal{A}}^\parallel(U)$. One additional linearly independent eigenvector has the form $(-\beta_2,\beta_1,0,0,0,0)$ and has zero eigenvalue.\footnote{Note that $\beta_1$ and $\beta_2$ cannot vanish simultaneously.} For $\tilde{\mathcal{A}}^\parallel(U)$ to be diagonalizable, there must exist a sixth linearly independent eigenvector of the form $(0,0,0,\vec{w})$ with $\vec{w}$ a nontrivial element of the kernel of $Q^\parallel(U_0)$. Therefore, one needs
\begin{align}
0 &= \det Q^\parallel(U_0) = 
\left(1-\frac {1}{d}\right)\begin{vmatrix}
    -1 & \delta_1  & 0 \\
    \alpha_4 & \delta_2 & \alpha_5 \\
    A_0 & 0 & -\frac{1}{d}
\end{vmatrix}
\nonumber\\
 &= \frac{1}{d}\left(1-\frac {1}{d}\right)\left[ (\alpha_4 + dA_0\alpha_5)\delta_1 + \delta_2 \right],
\end{align}
from which it follows that $\delta_2 = -\delta_1\left(dA_0\alpha_5 + \alpha_4\right)$. Together with the conditions~(\ref{Eq:deltaChoice1}) this leads to
\begin{equation}
\delta_1 = \frac{2\beta_3\eta}{\beta_1 - a\beta_2},\quad
\delta_2 = -\frac{2a\beta_3\eta}{\beta_1 - a\beta_2},\quad
\delta_3 = -2\eta,
\end{equation}
with $a := \alpha_4 + 2(d-1)\alpha_5\eta$. Of course, one should make sure that $\beta_1-a\beta_2\neq 0$; however for $\beta_2=0$ this is automatically satisfied. In this case, the above expressions simplify to
\begin{equation}
\delta_1 = -\frac{2\eta}{\kappa},\quad
\delta_2 = \frac{2a\eta}{\kappa},\quad
\delta_3 = -2\eta,
\label{Eq:delta}
\end{equation}
and
\begin{align}
\beta_1 = -\frac{k_B T}{e},
\quad 
\beta_2 = 0, \quad
\beta_3 = \frac{k_B T}{e}\frac{1}{\kappa}.
\label{Eq:betaCoeff1}
\end{align}

Summarizing what we have achieved so far, we have shown that the principal symbol $\mathcal{A}(k,U)$ is diagonalizable and has only real eigenvalues, provided that Eq.~(\ref{Eq:Gamma2}) and the inequality~(\ref{Eq:FinalHypoCausalBound}) hold and the coefficients $\delta_i$ are chosen according to Eq.~(\ref{Eq:delta}).

\subsection{Construction of the smooth symmetrizer}

Since the transformed symbol in Eq.~(\ref{Eq:SymbolScalarVecTensorForm}) has the block structure
\begin{equation}
\tilde{\mathcal{A}}(U) = \begin{pmatrix}
  0 & 0 & 0 \\
  0 & \tilde{\mathcal{A}}^\parallel(U) & 0 \\
  0 & 0 & \tilde{\mathcal{A}}^\perp(U)
\end{pmatrix},
\end{equation}
it is sufficient to find smooth symmetrizers $\tilde{H}^\parallel(U)$ and $\tilde{H}^\perp(U)$ such that $\tilde{H}^\parallel(U)\tilde{\mathcal{A}}^\parallel(U)$ and $\tilde{H}^\perp(U)\tilde{\mathcal{A}}^\perp(U)$ are symmetric. Indeed, if such symmetrizers are found, then it is simple to verify that
\begin{equation}
H(k,U) := T(k)^T\begin{pmatrix}
  I & 0 & 0 \\
  0 & \tilde{H}^\parallel(U) & 0 \\
  0 & 0 & \tilde{H}^\perp(U)
\end{pmatrix} T(k),
\end{equation}
is a symmetrizer for $\mathcal{A}(k,U)$.

Using the spectral decomposition one finds
\begin{equation}
\tilde{\mathcal{A}}^\parallel(U) = \sum\limits_{j=1}^5 \lambda_j P_j,
\end{equation}
where $\lambda_j$ refer to the five distinct eigenvalues $0$, $\pm\sqrt{\mu_1}$, $\pm\sqrt{\mu_2}$ of $\tilde{\mathcal{A}}^\parallel(U)$ and $P_j$ denote the corresponding eigenprojectors which satisfy $P_j P_k = \delta_{jk} P_j$ and $\sum_j P_j = I$. 
Since the matrix coefficients are smooth functions of the temperature $T$ only, classical results (see, for instance~\cite{Kato-Book}) imply that $\lambda_j$ and $P_j$ depend smoothly on $T$ as long as the eigenvalues do not cross (i.e. as long as they have constant multiplicity). Therefore, the symmetric positive definite matrix-value function of $T$
\begin{equation}
\tilde{H}^\parallel(U) := \sum\limits_{j=1}^5 P_j^T P_j
\end{equation}
depends smoothly on $T$. The construction for $\tilde{H}^\perp(U)$ is analogous.

\section{Propagation of the auxiliary constraints}
\label{Sec:ConstraintPropagation}

As we explained in Sec.~\ref{Sec:FOQ} the first-order system~(\ref{Eq:FOndot}--\ref{Eq:FOEdot}) is only equivalent to the original system~(\ref{Eq:ndot}--\ref{Eq:qdot}) when the constraints $C_\mu^{(\mathcal{N})} = 0$, $C_\mu^{(\mathcal{T})} = 0$, and $C_{\mu\nu}^{(B)} = 0$ are satisfied. In this section we prove that, provided the conditions (\ref{Eq:Gamma2}) and (\ref{Eq:delta}) hold, solutions of the first-order system automatically obey the constraints at all times if they do so at the initial time. In order to show this, we prove that, as a consequence of Eqs.~(\ref{Eq:FOndot}--\ref{Eq:FOEdot}), the constraint fields $C_\mu^{(\mathcal{N})}$, $C_\mu^{(\mathcal{T})}$ and $C_{\mu\nu}^{(B)}$ obey a homogenous system of evolution equations whose unique solution with trivial initial data is zero. This is achieved by embedding this evolution system into a larger one which is shown to be symmetric hyperbolic.

As a first step, we observe that the identities~(\ref{Eq:CNdot},\ref{Eq:CBdot}) together with the definitions~(\ref{Eq:Pialpha}--\ref{Eq:PiE}) and Eqs.~(\ref{Eq:FOndot}--\ref{Eq:FOmathcalTdot},\ref{Eq:FOBdot}) imply that
\begin{align}
\dot{C}^{(\mathcal{N})}_\mu &= 
 -\mathcal{N}^\nu C^{(B)}_{\mu\nu} 
 - k_\mu{}^\nu C_\nu^{(\mathcal{N})} + \delta_1 Z_\mu,
 \label{Eq:dotCN}\\
\dot{C}^{(\mathcal{T})}_\mu &=  -\mathcal{T}^\nu C^{(B)}_{\mu\nu} 
 - k_\mu{}^\nu C_\nu^{(\mathcal{T})} 
 + \Pi^{(\alpha)} C^{(\mathcal{T})}_\mu + \delta_2 Z_\mu,
\label{Eq:dotCT}\\
\dot{C}^{(B)}_{\mu\nu} &= -B^\alpha{}_\nu C^{(B)}_{\mu\alpha} - k_\mu{}^\alpha C^{(B)}_{\alpha\nu}
+C_\mu^{(\mathcal{T})} \left( \Pi^{(\beta)}_\nu + \Pi^{(E)}_\nu \right)
\nonumber\\
 &- \beta_1 Z^{(\mathcal{N})}_{\mu\nu}
 -\beta_2 Z^{(\mathcal{T})}_{\mu\nu},
\label{Eq:dotCB}
\end{align}
where we recall the definitions of  $Z^{(\mathcal{N})}_{\mu\nu}$, $Z^{(\mathcal{T})}_{\mu\nu}$, and $Z_\mu$ in Eqs.~(\ref{Eq:ZmathcalN},\ref{Eq:ZmathcalT},\ref{Eq:Z}). Although when using these definitions, Eqs.~(\ref{Eq:dotCN}--\ref{Eq:dotCB}) already yield an homogenous first-order evolution system for $C^{(\mathcal{N})}_\mu$, $C^{(\mathcal{T})}_\mu$ and $C^{(B)}_{\mu\nu}$, it fails to be strongly hyperbolic. For this reason, in a second step we enlarge the system by treating the $Z$-variables as independent and by deriving evolution equations for them.

In order to do so, we start with the definition~(\ref{Eq:ZmathcalN}) of $Z^{(\mathcal{N})}_{\mu\nu}$ and take a dot on both sides of this equation. Using the commutator identity~(\ref{Eq:Commutator1}) in App. \ref{App:Commutators} and Eqs.~(\ref{Eq:dotCN}) and (\ref{Eq:dotCB}) one first obtains
\begin{align}
\dot{Z}^{(\mathcal{N})}_{\mu\nu}
 &= -\mathcal{N}^\beta D_{[\mu}  C_{\nu]\beta}^{(B)} 
 + k_{[\mu}{}^\alpha D_{\nu]} C^{(\mathcal{N})}_\alpha 
 - k_{[\mu|\alpha|} D^\alpha C^{(\mathcal{N})}_{\nu]}
\nonumber\\
 &+ \delta_1 D_{[\mu} Z_{\nu]} + l.o.,
\label{Eq:dotZN}
\end{align}
where here and from now on ``$l.o.$" denotes terms which depend linearly on the constraint fields $C_\mu^{(\mathcal{N})}$, $C_\mu^{(\mathcal{T})}$, $C_{\mu\nu}^{(B)}$, $Z^{(\mathcal{N})}_{\mu\nu}$, $Z^{(\mathcal{T})}_{\mu\nu}$, and $Z^{(B)}_{\mu\nu\alpha}$ but not on their derivatives. Since
\begin{align}
k_{[\mu}{}^\alpha D_{\nu]} C^{(\mathcal{N})}_\alpha 
 & - k_{[\mu|\alpha|} D^\alpha C^{(\mathcal{N})}_{\nu]}
\nonumber\\
& = k_{\mu}{}^{\alpha}D_{[\nu}C_{\alpha]}^{(\mathcal{N})}-k_{\nu}{}^{\alpha}D_{[\mu}C_{\alpha]}^{(\mathcal{N})},
\end{align}
one can express the derivatives of the fields $C_{\mu\nu}^{(B)}$ and $C_\mu^{(\mathcal{N})}$ that appear on the first line of Eq.~(\ref{Eq:dotZN}) in terms of the (undifferentiated) fields $Z^{(B)}_{\mu\nu\alpha}$ and $Z^{(\mathcal{N})}_{\mu\nu}$, leaving only the antisymmetric derivative of the field $Z_\nu$ in the principal part, i.e. $\dot{Z}^{(\mathcal{N})}_{\mu\nu} = \delta_1 D_{[\mu} Z_{\nu]} + l.o.$ On the other hand, using the definition of $Z_\mu$ in Eq.~(\ref{Eq:Z}) and the commutator relations~(\ref{Eq:CommutatorBis0},\ref{Eq:CommutatorBis2}) one finds
\begin{equation}
D_{[\mu} Z_{\nu]} = D^\alpha Z^{(B)}_{\mu\nu\alpha} + l.o.,
\end{equation}
where again we have replaced antisymmetric derivatives of $C_{\mu\nu}^{(B)}$ with $Z^{(B)}_{\mu\nu\alpha}$. A similar calculation reveals that $\dot{Z}^{(\mathcal{T})}_{\mu\nu} = \delta_2 D^\alpha Z^{(B)}_{\mu\nu\alpha} + l.o.$ To obtain the evolution equation for $Z^{(B)}_{\mu\nu\alpha}$, we dot both sides of Eq.~(\ref{Eq:ZBN}). Using the identity~(\ref{Eq:Commutator2}) and Eq.~(\ref{Eq:dotCB}) we first obtain $\dot{Z}^{(B)}_{\mu\nu\alpha} = -\beta_1 D_{[\mu} Z^{(\mathcal{N})}_{\nu]\alpha} -\beta_2 D_{[\mu} Z^{(\mathcal{T})}_{\nu]\alpha} + l.o.$ Then, we use the commutator identity~(\ref{Eq:CommutatorBis1}) to obtain
\begin{equation}
D_{[\mu}Z^{(\mathcal N)}_{\nu]\alpha}	=\frac{1}{2}D_{\alpha}Z^{(\mathcal N)}_{\nu\mu}+l.o.,
\end{equation}
and similarly, $D_{[\mu}Z^{(\mathcal T)}_{\nu]\alpha} = -\frac{1}{2}D_{\alpha} Z^{(\mathcal T)}_{\mu\nu} + l.o$. Therefore, the evolution equation for $Z^{(B)}_{\mu\nu\alpha}$ has the form $\dot{Z}^{(B)}_{\mu\nu\alpha} = \frac{\beta_1}{2}D_{\alpha} Z^{(\mathcal N)}_{\mu\nu} + \frac{\beta_2}{2}D_{\alpha} Z^{(\mathcal T)}_{\mu\nu} + l.o.$ To simplify the final form of the system, we replace $Z^{(\mathcal{N})}_{\mu\nu}$ and $Z^{(\mathcal{T})}_{\mu\nu}$ with the new fields
\begin{align}
\mathcal{Z}_{\mu\nu} &:= \delta_2 Z_{\mu\nu }^{(\mathcal{N})} - \delta_1 Z_{\mu\nu}^{(\mathcal{T})},
\\
W_{\mu\nu} &:= \beta_1 Z_{\mu\nu}^{(\mathcal{N})} + \beta_2 Z_{\mu\nu}^{(\mathcal{T})}.
\end{align}
Note that the determinant of the transformation between $(Z^{(\mathcal{N})}_{\mu\nu},Z^{(\mathcal{T})}_{\mu\nu})$ and $(\mathcal{Z}_{\mu\nu},W_{\mu\nu})$ is equal to $s := \beta_1\delta_1 + \beta_2\delta_2$ which is positive according to Eqs.~(\ref{Eq:delta},\ref{Eq:betaCoeff1}). In terms of the fields $V := \left( C_\mu^{(\mathcal{N})},C_\mu^{(\mathcal{T})},C_{\mu\nu}^{(B)},\mathcal{Z}_{\mu\nu}, W_{\mu\nu},Z^{(B)}_{\mu\nu\alpha} \right)$ the propagation of the constraints is governed by a first-order evolution system of the form
\begin{equation}
\dot{V} = \mathcal{A}_c^\mu(U_0) D_\mu V + \mathcal{B}(U,DU) V,
\label{Eq:ConsPropSys}
\end{equation}
whose principal symbol $\mathcal{A}_c(U_0,k) := \mathcal{A}_c^\mu(U_0) k_\mu$ is given by
\begin{equation}
\mathcal{A}_c(U_0,k) V
= \begin{pmatrix}
 0 \\
 0 \\
 0 \\
 0 \\
 s k^\alpha Z^{(B)}_{\mu\nu\alpha}\\
 \frac{1}{2}k_\alpha W_{\mu\nu}
\end{pmatrix},
\end{equation}
and $\mathcal{B}$ is a matrix-valued function of the state vector $U$ and its derivatives.

As a final step, we show that the system (\ref{Eq:ConsPropSys}) is symmetric hyperbolic. Indeed, it is simple to verify that the symmetric, positive definite matrix
\begin{equation}
H_c(U_0) := 
\begin{pmatrix}
1 & 0 & 0 & 0 & 0 & 0 \\
0 & 1 & 0 & 0 & 0 & 0 \\
0 & 0 & 1 & 0 & 0 & 0 \\
0 & 0 & 0 & 1 & 0 & 0 \\
0 & 0 & 0 & 0 & 1 & 0 \\
0 & 0 & 0 & 0 & 0 & 2s \\
\end{pmatrix}
\end{equation}
is a symmetrizer for $\mathcal{A}_c(U_0,k)$, that is, $H_c(U_0)\mathcal{A}_c(U_0,k)$ is symmetric for all $U_0$ and $k$. Since $H_c$ is independent of $k$ this implies that the constraint propagation system~(\ref{Eq:ConsPropSys}) is symmetric hyperbolic. In particular, $V = 0$ is the unique solution with trivial initial data, and hence the constraints propagate as desired.

\section{Conclusions}
\label{Sec:Conclusions}

In this work and the companion letter~\cite{aGjSoS2024a} we presented and analyzed a novel theory for relativistic dissipative fluids in the spirit of BDNK theories. The particular frame our theory is based on, which we refer to as the trace-fixed particle frame, is formally motivated by a microscopic theory whose details will be discussed elsewhere. This fact contributes to the physical interpretation of the general first-order theories and may shed light on the problem of dissipative phenomena in relativistic systems. 

We would like to emphasize that, once the coefficients $\Gamma_1$ and $\Gamma_2$ have been fixed according to Eqs.~(\ref{Eq:Gamma1}) and (\ref{Eq:Gamma2}), strong hyperbolicity and causality only require the fulfillment of the single inequality~(\ref{Eq:FinalHypoCausalBound}). Furthermore, this condition is independent of the only arbitrary parameter $\Lambda_0$ in our system, whose sole purpose is to achieve the stability property. Therefore, we believe our theory constitutes a drastic simplification over previous results which usually contain an ample set of parameters and inequalities.

The main result of this article consists in proving that the proposed fluid theory is governed by strongly hyperbolic and causal evolution  equations in the full nonlinear regime. Moreover, as shown in App.~\ref{App:Entropy}, it is compatible with the second law of thermodynamics within its domain of validity. These properties, in conjunction with the stability of the equilibrium configuration shown in~\cite{aGjSoS2024a} render this theory a promising candidate for describing nonequilibrium processes. In particular, our approach provides transport equations which lead to a well-posed Cauchy problem (see Theorem~1), which is key to the predictive power of the theory and fundamental for numerical simulations.

There are some important open questions that have been left unanswered in this article and that will be analyzed in future work. A pressing question is whether our main theorem on the local well posedness can be extended to the full Einstein-fluid equations in which the spacetime metric is evolved along with the fluid instead of being held fixed. This is important in view of physical scenarios in which the self-gravity of the fluid must be included. Another relevant extension of our work relies in considering a fluid mixture consisting of several charged species and analyzing the resulting Maxwell-fluid system.


\acknowledgments

We thank G. Chac\'on-Acosta, L. Lehner, A. R. M\'endez, O.~Reula, and T. Zannias for fruitful comments and discussions. We also thank L. Lehner for comments on a previous version of this manuscript. O.S. was partially supported by CIC Grant No.~18315 to Universidad Michoacana and by CONAHCyT Network Project No.~376127 ``Sombras, lentes y ondas gravitatorias generadas por objetos compactos astrof\'isicos". FS was supported by a CONAHCyT postdoctoral fellowship.

\appendix

\section{First-order theories and invariant definition of transport coefficients}
\label{App:FirstOrderDissipativeFluids}

In this appendix we review the method discussed by Kovtun in~\cite{pK19} to define transport coefficients which are invariant with respect to first-order changes of frames. In addition, we refine Kovtun's method to include changes of representation~\cite{JNET24} which are due to the use of the equations of motion, and construct transport coefficients which are both frame and representation invariant.

For this, we start with the general decomposition of the current density and stress-energy tensor,
\begin{eqnarray}
J^\mu &=& \mathcal{N} u^\mu + \mathcal{J}^\mu,\\
T^{\mu\nu} &=& \mathcal{E} u^\mu u^\nu + \mathcal{P}\Delta^{\mu\nu} + 2 u^{(\mu}\mathcal{Q}^{\nu)} + \mathcal{T}^{\mu\nu},
\label{Eq:TmunuKovtun}
\end{eqnarray}
with respect to a future-directed timelike vector field $u^\mu$, normalized such that $u^\mu u_\mu = -1$. As before, $\Delta^{\mu\nu} = g^{\mu\nu} + u^\mu u^\nu$ and $\mathcal{J}^\mu$, $\mathcal{Q}^\mu$, $\mathcal{T}^{\mu\nu}$ are orthogonal to $u^\mu$, and $\mathcal{T}^{\mu\nu}$ is symmetric and trace-free. Note that we are now working in an arbitrary frame, where $u^\mu$ is not necessarily aligned with the mean particle flow. Following~\cite{pK19}, we consider the most general constitutive relations that can be constructed from first-order gradients of the fundamental fields $(n,T,u^\mu)$:
\begin{align}
\mathcal{N} &= n 
 + \nu_1\frac{\dot{n}}{n} + \nu_2\frac{\dot{T}}{T}
 + \nu_3\theta + \mathcal{O}(\partial^2),
\label{Eq:ConstitutiveN}\\
\mathcal{E} &= n e(n,T) + \varepsilon_1\frac{\dot{n}}{n} 
 + \varepsilon_2\frac{\dot{T}}{T} + \varepsilon_3\theta 
 + \mathcal{O}(\partial^2),
\label{Eq:ConstitutiveE}\\
\mathcal{P} &= p(n,T) 
 + \pi_1\frac{\dot{n}}{n} + \pi_2\frac{\dot{T}}{T} + \pi_3\theta + \mathcal{O}(\partial^2),
\label{Eq:ConstitutiveP}\\
\mathcal{J}^\mu &= \gamma_1\frac{D^\mu n}{n}
 + \gamma_2\frac{D^\mu T}{T} + \gamma_3 a^\mu 
 + \gamma_4\frac{q}{h} F^{\mu\nu}u_\nu
 + \mathcal{O}(\partial^2),
\label{Eq:ConstitutiveJ}\\
\mathcal{Q}^\mu &= \kappa_1\frac{D^\mu n}{n}
 + \kappa_2\frac{D^\mu T}{T} + \kappa_3 a^\mu 
 + \kappa_4\frac{q}{h} F^{\mu\nu}u_\nu
 + \mathcal{O}(\partial^2),
\label{Eq:ConstitutiveQ}\\
\mathcal{T}^{\mu\nu} &= -2\eta\sigma^{\mu\nu}
 + \mathcal{O}(\partial^2),
\label{Eq:ConstitutiveT}
\end{align}
where $\mathcal{O}(\partial^n)$ denotes terms that are at least $n$th-order in the gradients of the fields. Here, the $18$ transport coefficients $\nu_i$ $\varepsilon_i$, $\pi_i$ ($i=1,2,3$), $\gamma_j$, $\kappa_j$ ($j=1,2,3,4$), and $\eta$ are functions of $(n,T)$ only. Note that in contrast to Ref.~\cite{pK19}, here we take into account the (background) electromagnetic field $F^{\mu\nu}$, which is assumed to be first order in the gradients. This leads to the additional two transport coefficients $\gamma_4$ and $\kappa_4$.

It is important to realize that $\nu_i$ $\varepsilon_i$, $\pi_i$, $\gamma_j$, $\kappa_j$ are not invariant. They depend on the choice of the frame and they can be altered by using the equations of motion resulting in a change of representation. Furthermore, consistency with the second law of thermodynamics implies constraints for the coefficients $\gamma_j$ and $\kappa_j$. In the following, we discuss each of these properties and show how they can be used to introduce three invariant quantities which will be identified with the shear and bulk viscosity coefficients $\eta$ and $\zeta$ respectively, and the thermal conductivity $\kappa$.

\subsection{First-order changes of frame}

The determination of $(n,T,u^\mu)$ from a given, off-equilibrium configuration is not unique. A specific choice is referred to as a ``frame" in the literature. In the following, we consider a first-order change of the frame, that is, a transformation of the form
\begin{eqnarray}
n &\mapsto& n' := n + \delta n,
\label{Eq:ChangeOfFramen}\\
T &\mapsto& T' := T + \delta T,
\label{Eq:ChangeOfFrameT}\\
u^\mu &\mapsto& {u'}^\mu := u^\mu + \delta u^\mu,
\label{Eq:ChangeOfFrameu}
\end{eqnarray}
where the quantities $\delta n$, $\delta T$, and $\delta u^\mu$ are of order $\mathcal{O}(\partial^1)$. Note that $u'_\mu {u'}^\mu = -1$ implies that
\begin{equation}
u_\mu\delta u^\mu = \mathcal{O}(\partial^2).
\end{equation}
Under a first-order change of frame, it is not difficult to show that the quantities $\mathcal{N}$, $\mathcal{E}$, $\mathcal{P}$, $\mathcal{J}^\mu$, $\mathcal{Q}^\mu$ and $\mathcal{T}^{\mu\nu}$ transform according to
\begin{eqnarray}
\mathcal{N}' &=& \mathcal{N} + \mathcal{O}(\partial^2),
\\
\mathcal{E}' &=& \mathcal{E} + \mathcal{O}(\partial^2),
\\
\mathcal{P}' &=& \mathcal{P} + \mathcal{O}(\partial^2),
\\
{\mathcal{J}'}^\mu &=& \mathcal{J}^\mu - n\delta u^\mu 
 + \mathcal{O}(\partial^2),
\\
{\mathcal{Q}'}^\mu &=& \mathcal{Q}^\mu - nh\delta u^\mu 
 + \mathcal{O}(\partial^2),
\\
{\mathcal{T}'}^{\mu\nu} &=& \mathcal{T}^{\mu\nu}
 + \mathcal{O}(\partial^2).
\end{eqnarray}
Expanding the primed quantities in the same fashion as in Eqs.~(\ref{Eq:ConstitutiveN}--\ref{Eq:ConstitutiveT}) and assuming
\begin{align}
\delta n &= 
\alpha_1\frac{\dot{n}}{n} + \alpha_2\frac{\dot{T}}{T}
 + \alpha_3\theta + \mathcal{O}(\partial^2),
\\
\delta T &= 
\beta_1\frac{\dot{n}}{n} + \beta_2\frac{\dot{T}}{T}
 + \beta_3\theta + \mathcal{O}(\partial^2),
\\
\delta u^\mu &= \mu_1\frac{D^\mu n}{n}
 + \mu_2\frac{D^\mu T}{T} + \mu_3 a^\mu 
 + \mu_4 \frac{q}{h} F^{\mu\nu} u_\nu
 + \mathcal{O}(\partial^2),
\end{align}
with the ten free coefficients $\alpha_i$, $\beta_i$ and $\mu_j$ depending only on $(n,T)$, one finds the following relation between the primed and unprimed transport coefficients:
\begin{align}
\nu_i' &= \nu_i - \alpha_i,
\label{Eq:nuprime}\\
\varepsilon_i' &= \varepsilon_i 
 - \left. \frac{\partial(ne)}{\partial n} \right|_T\alpha_i
 - n c_v\beta_i,\qquad
 c_v := \left. \frac{\partial e}{\partial T} \right|_n,
\\
\pi_i' &= \pi_i
 - \left. \frac{\partial p}{\partial n} \right|_T\alpha_i
 - \left. \frac{\partial p}{\partial T} \right|_n\beta_i,
\\
\gamma'_j &= \gamma_j - n\mu_j,
\\
\kappa'_j &= \kappa_j - nh\mu_j,
\\
\eta' &= \eta.
\label{Eq:etaprime}
\end{align}
Therefore, whereas $\eta$ is frame independent, the remaining coefficients $\nu_i$, $\varepsilon_i$, $\pi_i$, $\gamma_j$ and $\kappa_j$ are frame-dependent. However, one can construct eight quantities which are invariant with respect to the transformations given by Eqs. (\ref{Eq:ChangeOfFramen}--\ref{Eq:ChangeOfFrameu}): $\eta$ and
\begin{align}
f_i &:=\pi_i 
 - \frac{1}{n c_v}\left. \frac{\partial p}{\partial T} \right|_n \varepsilon_i \nonumber \\
 & + \left[\frac{1}{n c_v}\left. \frac{\partial p}{\partial T} \right|_n\left. \frac{\partial(ne)}{\partial n} \right|_T -  \left. \frac{\partial p}{\partial n} \right|_T  \right]\nu_i,
\label{Eq:FrameInvariantf}\\
\ell_j &:= \gamma_j - \frac{1}{h}\kappa_j,
\label{Eq:FrameInvariantl}
\end{align}
which can thus be evaluated in any frame.

\subsection{Changes of representation}

As mentioned before, there is another freedom in the transport coefficients that should be taken into account. Indeed, the equations of motion~(\ref{eq:euler}) imply that
\begin{align}
\frac{\dot{n}}{n} + \theta =\mathcal{O}(\partial^2),
\label{Eq:EulerTrunc1}\\
\frac{\dot{T}}{T} + \frac{k_B}{c_v}\theta = \mathcal{O}(\partial^2),
\label{Eq:EulerTrunc2}\\
a^\mu + \frac{1}{n h} D^\mu p - \frac{q}{h} F^{\mu\nu}u_\nu
= \mathcal{O}(\partial^2),
\label{Eq:EulerTrunc3}
\end{align}
where from now on we assume the ideal gas equation of state $p(n,T) = n k_B T$ for simplicity. Therefore, one can modify, for instance, Eq.~(\ref{Eq:ConstitutiveP}) to
\begin{align}
\mathcal{P} &= p(n,T) 
 + \pi_1\frac{\dot{n}}{n} + \pi_2\frac{\dot{T}}{T} + \pi_3\theta \nonumber\\ 
 &+ \Omega_\pi\left( \frac{\dot{n}}{n} + \theta \right)
 + \chi_\pi\left( \frac{\dot{T}}{T} + \frac{k_B}{c_v}\theta \right) 
 + \mathcal{O}(\partial^2),
\end{align}
with some arbitrary coefficients $\Omega_\pi$ and $\chi_\pi$ depending only on $(n,T)$. This yields the same structure for $\mathcal P$  as the one in Eq.~(\ref{Eq:ConstitutiveP}) but with the modified coefficients
\begin{eqnarray}
\hat{\pi}_1 &=& \pi_1 + \Omega_\pi,\\
\hat{\pi}_2 &=& \pi_2 + \chi_\pi,\\
\hat{\pi}_3 &=& \pi_3 + \Omega_\pi + \frac{k_B}{c_v}\chi_\pi,
\end{eqnarray}
and hence both sets $(\pi_1,\pi_2,\pi_3)$ and $(\hat{\pi}_1,\hat{\pi}_2,\hat{\pi}_3)$ should be considered to be equivalent to each other. Similar relations hold for the remaining scalar coefficients $\nu_i$ and $\varepsilon_i$.

Likewise, one can modify Eq.~(\ref{Eq:ConstitutiveQ}) to
\begin{align}
\mathcal{Q}^\mu &= \kappa_1\frac{D^\mu n}{n}
 + \kappa_2\frac{D^\mu T}{T} + \kappa_3 a^\mu 
 + \kappa_4\frac{q}{h} F^{\mu\nu} u_\nu \nonumber\\ 
 &+ \xi_Q\left( a^\mu + \frac{1}{n h} D^\mu p 
 - \frac{q}{h} F^{\mu\nu} u_\nu \right)
 + \mathcal{O}(\partial^2),
\end{align}
which is equivalent to transforming the coefficients $(\kappa_1,\kappa_2,\kappa_3,\kappa_4)$ to
\begin{eqnarray}
\hat{\kappa}_1 &=& \kappa_1 + \frac{k_B T}{h} \xi_Q,\\
\hat{\kappa}_2 &=& \kappa_2 + \frac{k_B T}{h} \xi_Q,\\
\hat{\kappa}_3 &=& \kappa_3 + \xi_Q,\\
\hat{\kappa}_4 &=& \kappa_4 - \xi_Q.
\end{eqnarray}
Again, both sets should be considered to be equivalent, and similar relations hold for the coefficients $\gamma_j$.

Taking this into account, one can construct invariant quantities in a similar fashion as for the frame transformation. Indeed the following combinations
\begin{align}
\nu_1 + \frac{k_B}{c_v}\nu_2 - \nu_3,\,\,
\varepsilon_1 + \frac{k_B}{c_v}\varepsilon_2 - \varepsilon_3,\,\,
\pi_1 + \frac{k_B}{c_v}\pi_2 - \pi_3,
\end{align}
and
\begin{align}
&& \gamma_1 - \frac{k_B T}{h}\gamma_3,\quad
\gamma_2 - \frac{k_B T}{h} \gamma_3,\quad
\gamma_3 + \gamma_4,
\\
&& \kappa_1 - \frac{k_B T}{h}\kappa_3,\quad
\kappa_2 - \frac{k_B T}{h}\kappa_3,\quad
\kappa_3 + \kappa_4,
\end{align}
are invariant with respect to changes of representation (the hatted transformations) and thus are independent of the way the equations of motion are used to rewrite the constitutive relations.

Combining these invariants with the frame-independent quantities $f_i$ and $\ell_j$ one finds the following five quantities
\begin{align}
&& \eta, \label{inveta}
\\
&& \zeta := f_1 + \frac{k_B}{c_v} f_2 - f_3,
\\
&& \lambda_1 := \ell_1 - \frac{k_B T}{h}\ell_3,
\label{invlambda1}\\
&& \lambda_2 := \ell_2 - \frac{k_B T}{h}\ell_3,
\label{invlambda2}\\
&& \lambda_3 := \ell_3 + \ell_4,
\label{invlambda3}
\end{align}
which are invariant with respect to both first-order frame transformations and changes of representation. As we will see shortly, $\zeta$ is the bulk viscosity parameter and $\lambda_1$ is proportional to the thermal conductivity.

\subsection{Thermodynamic constraints}

Up to now, the number of coefficients involved in the general constitutive relations was reduced from $18$ to $5$ by means of considering two freedoms that are allowed within the first-order scheme: a change of frame given by Eqs.~(\ref{Eq:ChangeOfFramen}-\ref{Eq:ChangeOfFrameu}) and the addition of multiples of the left-hand sides of Eqs.~(\ref{Eq:EulerTrunc1}-\ref{Eq:EulerTrunc3}). These transformations led to five invariant quantities by means of enforcing the fluxes $T^{\alpha\beta}$ and $J^\alpha$ to remain unaltered in the transformation to the corresponding first order. 
However, there is one final consideration that allows the further reduction of the number of invariant transport coefficients from $5$ to $3$. This final step is rooted in the laws of thermodynamics, in particular the requirement of a non-negative entropy production. In Appendix~\ref{App:Entropy} the expression for such quantity is derived in a frame-independent way, which leads to the constraints for $\ell_i$ given by in Eq. (\ref{contstraintsells}). Notice that such relations, when written in terms of $\lambda_i$ (see Eq.~(\ref{invlambda1}-\ref{invlambda3})) finally lead to \begin{equation}
\frac{e}{k_B T}\lambda_1 + \lambda_2 = 0,\qquad
\frac{h}{k_B T}\lambda_1 + \lambda_3 = 0.
\label{Eq:ThermoConstraints}
\end{equation}
This indicates that although Eqs.~(\ref{inveta}-\ref{invlambda3}) feature five invariant quantities, the coefficients $\lambda_i$ are not independent, which finally reduces the number of transport coefficients to $3$.

\subsection{Summary}

In order to shed light on the physical meaning of the invariant coefficients defined so far, it is instructive to write down the constitutive relations in the Eckart frame, for which $\nu_i = \varepsilon_i = 0$ and $\gamma_j = 0$ (note that this frame can always be attained by means of the transformation~(\ref{Eq:nuprime}--\ref{Eq:etaprime})), such that $f_i = \pi_i$ and $\ell_j = -\kappa_j/h$. Furthermore, one can use the hatted transformations to achieve $\pi_1 = \pi_2 = 0$ and $\kappa_1 = 0$. Then, the constraints~(\ref{Eq:ThermoConstraints}) imply that $\kappa_4 = 0$ and $\kappa_2 = \kappa_3$, and it follows that the invariants reduce to $\zeta = -\pi_3$, and (assuming again $p = n k_B T$) 
\begin{equation}
\lambda_1 = \frac{k_B T}{h^2}\kappa_2,\qquad
\lambda_2 = -\frac{e}{h^2}\kappa_2,\qquad
\lambda_3 = -\frac{\kappa_2}{h}.
\end{equation}
In this frame, one thus has $J^\mu = n u^\mu$, $T^{\mu\nu} = n e u^\mu u^\nu + \mathcal{P}\Delta^{\mu\nu} + 2u^{(\mu}\mathcal{Q}^{\nu)} + \mathcal{T}^{\mu\nu}$ with the constitutive relations
\begin{eqnarray}
\mathcal{P} &=& p - \zeta\theta + \mathcal{O}(\partial^2),
\\
\mathcal{Q}^\mu &=& \frac{h^2\lambda_1}{k_B T}\left( \frac{D^\mu T}{T}
 + a^\mu \right) + \mathcal{O}(\partial^2),
\\
\mathcal{T}^{\mu\nu} &=& -2\eta\sigma^{\mu\nu}
 + \mathcal{O}(\partial^2).
\end{eqnarray}
From this it is clear that $\zeta$ is the bulk viscosity coefficient and $\kappa := -h^2\lambda_1/(k_B T)$ the thermal conductivity.

Summarizing, we end up with three independent invariant transport coefficients. In an arbitrary frame they are given by
\begin{align}
\eta && \hbox{(shear viscosity)},\\
\zeta = f_1 + \frac{k_B}{c_v} f_2 - f_3 &&
\hbox{(bulk viscosity)},\\
\kappa = -h\left( \ell_1 - \ell_2 \right) &&
\hbox{(thermal conductivity)},
\end{align}
where the invariants $f_1$, $f_2$, $f_3$, $\ell_1$ and $\ell_3$ are defined in Eqs.~(\ref{Eq:FrameInvariantf}, \ref{Eq:FrameInvariantl}). In a particle frame, for which $J^\mu = n u^\mu$, one has $\nu_i = 0$ and $\gamma_j = 0$, and allowing $\varepsilon_i\neq 0$ one has
\begin{align}
\zeta &= \left( \pi_1 - \frac{k_B}{c_v}\varepsilon_1 \right)
 + \frac{k_B}{c_v} \left( \pi_2 - \frac{k_B}{c_v}\varepsilon_2 \right)
 - \left( \pi_3 - \frac{k_B}{c_v}\varepsilon_3 \right),
\label{Eq:BulkPF}\\
\kappa &= \kappa_1 - \kappa_2.
\label{Eq:ThermalPF}
\end{align}

\section{Entropy current and entropy production}
\label{App:Entropy}

In this appendix we briefly review a definition of the entropy current which is first-order frame invariant and compute the corresponding entropy production. Furthermore, we derive the restrictions on the constitutive relations that are imposed by requiring that solutions of the equations of motion have non-negative entropy production. 

The entropy current can be defined as~\cite{Israel1979,Israel1987}
\begin{equation}
S^\alpha := \frac{1}{T}\left( p u^\alpha -  T^{\alpha\beta} u_\beta - \mu J^\alpha \right),
\label{Eq:EntropyFlux}
\end{equation}
where here $p$, $s$, and $\mu$ refer to the pressure, entropy per particle and the chemical potential (or Gibbs potential per particle),
\begin{equation}
\mu := h - T s = e + \frac{p}{n} - T s.
\label{Eq:GibbsPotential}
\end{equation}
Note that $p$, $s$, and $\mu$ are functions of the \emph{equilibrium quantities} $(n,T)$ and hence, $S^\alpha$ is frame-dependent in general. According to the first law of thermodynamics,
\begin{equation}
d\mu = \frac{dp}{n} - s dT.
\label{Eq:FirstLaw}
\end{equation}
In equilibrium Eq.~(\ref{Eq:EntropyFlux}) yields $S^\alpha = (p + n e - \mu n) u^\alpha/T = n s u^\alpha$, as expected. With respect to an arbitrary variation of an equilibrium state, using Eqs.~(\ref{Eq:GibbsPotential},\ref{Eq:FirstLaw}), one finds the following expression for the first variation $\delta S^\alpha$ of the entropy current:
\begin{equation}
\delta S^\alpha = -\frac{u_\beta}{T} \delta T^{\alpha\beta} - \frac{\mu}{T}\delta J^\alpha,
\end{equation}
which is the covariant Gibbs relation~\cite{Israel1979,Israel1987}. Specializing this relation to a first-order change of frame (see Eqs.~(\ref{Eq:ChangeOfFramen}--\ref{Eq:ChangeOfFrameu})), for which $\delta J^\alpha = 0$ and $\delta T^{\alpha\beta} = 0$, one finds $\delta S^\alpha = 0$ which implies that the entropy current in Eq.~(\ref{Eq:EntropyFlux}) is first-order frame invariant.

Next, we compute the entropy production. For simplicity we adopt a frame in which $J^\alpha = n u^\alpha$; however the final result will not depend on this election since $S^\alpha$ is first-order frame invariant. Applying the divergence on both sides of Eq.~(\ref{Eq:EntropyFlux}) and using the equations of motion~(\ref{eq:euler}) with $T^{\alpha\beta}$ given by Eq.~(\ref{Eq:TmunuKovtun}), we obtain
\begin{align}
T\nabla_\alpha S^\alpha &= -(\mathcal{E} - ne)\frac{\dot{T}}{T} - (\mathcal{P} - p)\theta
\nonumber\\
 &- \mathcal{Q}^\alpha\left( \frac{D_\alpha T}{T} + a_\alpha \right) - \mathcal{T}^{\alpha\beta}\sigma_{\alpha\beta}.
\label{Eq:EntropyProduction}
\end{align}
As observed in~\cite{pK19} the entropy production should be non-negative when evaluated on shell, i.e. for solutions satisfying the evolution equations. For this reason, we eliminate $\dot{T}/T$, $\dot{n}/n$, and $D^\alpha n/n$ in Eqs.~(\ref{Eq:EntropyProduction}) and the constitutive relations (\ref{Eq:ConstitutiveE},\ref{Eq:ConstitutiveP},\ref{Eq:ConstitutiveQ}) using the fact that the Euler equations are satisfied up to second-order corrections in the gradients (see Eqs.~(\ref{Eq:EulerTrunc1}--\ref{Eq:EulerTrunc3})). After some calculations one finds
\begin{align}
& T\nabla_\alpha S^\alpha = \zeta\theta^2 + (\kappa_1-\kappa_2) A^\alpha A_\alpha
 + 2\eta \sigma^{\alpha\beta}\sigma_{\alpha\beta}
\nonumber\\
 &+ \left( \frac{e\kappa_1}{k_B T} + \kappa_2 - \kappa_3 \right) a^\alpha A_\alpha
 - \frac{q}{h}\left( \frac{h \kappa_1}{k_B T} + \kappa_4 \right) E^\alpha A_\alpha
\nonumber\\
 &+ \mathcal{O}(\partial^3),
\end{align}
where we have abbreviated $A^\alpha := a^\alpha + D^\alpha T/T$, $E^\mu := F^{\mu\nu} u_\nu$ and where $\zeta$ is the bulk viscosity defined in Eq.~(\ref{Eq:BulkPF}). Non-negative entropy production requires 
\begin{equation}
\zeta\geq 0,\quad
\kappa_1 - \kappa_2\geq 0,\quad
\eta\geq 0,
\end{equation}
and the satisfaction of the two relations
\begin{equation}
\frac{e}{k_B T}\kappa_1 + \kappa_2 - \kappa_3 = 0,\quad
\frac{h}{k_B T}\kappa_1 + \kappa_4 = 0.
\label{contstraintskappas}
\end{equation}
The combination $\kappa_1-\kappa_2$ is the thermal conductivity $\kappa$ defined in Eq.~(\ref{Eq:ThermalPF}). The final result for the entropy production is
\begin{equation}
\nabla_\alpha S^\alpha = \frac{\zeta}{T}\theta^2 + \frac{\kappa}{T} A^\alpha A_\alpha
 + \frac{2\eta}{T} \sigma^{\alpha\beta}\sigma_{\alpha\beta}
 + \mathcal{O}(\partial^3),
\end{equation}
and is valid in an arbitrary frame. In terms of the first-order frame-invariant quantities $\ell_j$ defined in Eq.~(\ref{Eq:FrameInvariantl}), Eq.~(\ref{contstraintskappas}) yields
\begin{equation}
\frac{e}{k_B T}\ell_1 + \ell_2 - \ell_3 = 0,\quad
\frac{h}{k_B T}\ell_1 + \ell_4 = 0.
\label{contstraintsells}
\end{equation}
Finally, if the transport coefficients $\zeta$, $\kappa$ and $\eta$ are strictly positive, then the entropy production is zero only if
\begin{equation}
\theta = 0,\quad
\sigma_{\alpha\beta} = 0,\quad
a^\alpha + \frac{D^\alpha T}{T} = 0,
\end{equation}
which are the conditions for global equilibrium.

\section{Commutators}
\label{App:Commutators}

In this appendix we summarize some useful commutator identities that are needed throughout this article. The following relations hold on an arbitrary, sufficiently smooth, spacetime manifold $(M,g)$ of dimension $d+1$ with associated Levi-Civita connection $\nabla$ and Riemann tensor $R^\alpha{}_{\beta\mu\nu}$. As in the main text,
\begin{equation}
a^\mu := u^\nu\nabla_\nu u^\mu,\qquad
k_{\mu\nu} := \Delta_\mu{}^\alpha\nabla_\alpha u_\nu,
\end{equation}
denote the acceleration and spatial gradient of the unit timelike vector field $u^\mu$, respectively, where $\Delta_\mu{}^\nu := \delta_\mu{}^\nu + u_\mu u^\nu$ refers to the projector orthogonal to $u^\mu$. Note that $k_{\mu\nu}$ is symmetric only if $u^\mu$ is hypersurface orthogonal. In this case, $k_{\mu\nu}$ describes the second fundamental form associated with the surface orthogonal to $u^\mu$. For any covariant tensor field $T_{\mu_1\mu_2\ldots\mu_s}$ orthogonal to $u^\mu$ we define
\begin{align}
\dot{T}_{\mu_1\mu_2\ldots\mu_s} &:= \Delta_{\mu_1}{}^{\nu_1}\Delta_{\mu_2}{}^{\nu_2}\cdots\Delta_{\mu_s}{}^{\nu_s} u^\alpha\nabla_\alpha T_{\nu_1\nu_2\ldots\nu_s},
\\
D_\alpha T_{\mu_1\mu_2\ldots\mu_s}
 &:= \Delta_\alpha{}^\beta\Delta_{\mu_1}{}^{\nu_1}\cdots\Delta_{\mu_s}{}^{\nu_s} \nabla_\beta T_{\nu_1\nu_2\ldots\nu_s}.
\end{align}
With these definitions, one finds the following identities for smooth functions $f$, $1$-forms $X_\mu$, and tensor fields $T_{\mu\nu}$ (the latter two being orthogonal to $u^\mu$):
\begin{align}
(D_\mu f)\dot{} - D_\mu\dot{f} &= a_\mu\dot{f} - k_\mu{}^\nu D_\nu f,
\label{Eq:Commutator0}\\
(D_\mu X_\nu)\dot{} - D_\mu\dot{X}_\nu
 &= a_\mu\dot{X}_\nu - k_\mu{}^\beta D_\beta X_\nu + \mathcal{K}_{\mu\nu}{}^\beta X_\beta,
\label{Eq:Commutator1}\\
(D_\mu T_{\nu\alpha})\dot{} - D_\mu\dot{T}_{\nu\alpha}
 &= a_\mu\dot{T}_{\nu\alpha} -k_\mu{}^{\beta}D_{\beta} T_{\nu\alpha}
\nonumber\\
 &+ \mathcal{K}_{\mu\nu}{}^\beta T_{\beta\alpha}
 + \mathcal{K}_{\mu\alpha}{}^\beta T_{\nu\beta},
\label{Eq:Commutator2}
\end{align}
where we have introduced the tensor field
\begin{equation}
\mathcal{K}_{\mu\nu\beta} := k_{\mu\nu} a_\beta - k_{\mu\beta} a_\nu - \Delta_\nu{}^{\nu'}\Delta_\beta{}^{\beta'} R_{\nu'\beta'\mu\sigma} u^\sigma,
\end{equation}
which is antisymmetric in the last two indices.

Next, the antisymmetric spatial second derivatives of $f$, $X_\mu$, and $T_{\mu\nu}$ are given by
\begin{align}
D_{[\mu} D_{\nu]} f &= k_{[\mu\nu]}\dot{f},
\label{Eq:CommutatorBis0}\\
D_{[\mu} D_{\nu]} X_\alpha &= k_{[\mu\nu]}\dot{X}_\alpha - \frac{1}{2} \mathcal{H}_{\mu\nu\alpha}{}^\beta X_\beta,
\label{Eq:CommutatorBis1}\\
D_{[\mu} D_{\nu]} T_{\alpha\beta} &= k_{[\mu\nu]} \dot{T}_{\alpha\beta} - \frac{1}{2} \mathcal{H}_{\mu\nu\alpha}{}^\sigma T_{\sigma\beta}  
 - \frac{1}{2}\mathcal{H}_{\mu\nu\beta}{}^\sigma T_{\alpha\sigma},
\label{Eq:CommutatorBis2}
\end{align}
where we have set
\begin{equation}
\mathcal{H}_{\mu\nu\alpha\beta} := k_{\mu\alpha} k_{\nu\beta} - k_{\nu\alpha} k_{\mu\beta} - \Delta_\mu{}^{\mu'}\Delta_\nu{}^{\nu'}\Delta_\alpha{}^{\alpha'}\Delta_\beta{}^{\beta'} R_{\mu'\nu'\alpha'\beta'}.
\end{equation}
Notice that $D_{[\mu} D_{\nu]} f = 0$ only if $u^\mu$ is hypersurface orthogonal. In this case, $D_\mu$ is the induced covariant derivative along the surface $\Sigma$ orthogonal to $u^\mu$ and Eq.~(\ref{Eq:CommutatorBis1}) yields the Gauss equation
\begin{equation}
\Delta_\mu{}^{\mu'}\Delta_\nu{}^{\nu'}\Delta_\alpha{}^{\alpha'}\Delta_\beta{}^{\beta'} R_{\mu'\nu'\alpha'\beta'} = \bar{R}_{\mu\nu\alpha\beta} + k_{\mu\alpha} k_{\nu\beta} - k_{\nu\alpha} k_{\mu\beta},
\end{equation}
with $\bar{R}_{\mu\nu\alpha\beta}$ the intrinsic curvature of $\Sigma$. Contracting twice one also obtains the following equation, familiar from the $3+1$ decomposition in general relativity,
\begin{equation}
2G_{\mu\nu} u^\mu u^\nu = \bar{R} + k^2 - k^{\alpha\beta} k_{\alpha\beta},
\end{equation}
with $k$ the trace of $k_{\mu\nu}$ and $G_{\mu\nu}$ the Einstein tensor.

Finally, we also note the relations
\begin{align}
& \dot{k}_{\mu\nu} = \left(D_\mu + a_\mu\right)a_\nu - k_\mu{}^{\alpha} k_{\alpha\nu} - R_{\mu\alpha\nu\beta} u^\alpha u^\beta,
\label{Eq:Dotk}\\
& 2D_{[\mu} k_{\nu]\alpha} = \Delta_\mu{}^{\mu'}\Delta_\nu{}^{\nu'} R_{\mu'\nu'\alpha\beta} u^\beta + 2k_{[\mu\nu]} a_\alpha.
\label{Eq:Curlk}
\end{align}
Taking the trace, symmetric trace-free and antisymmetric parts of Eq.~(\ref{Eq:Dotk})  yields~\cite{Wald1984,Ellis2012}
\begin{align}
\dot{\theta} &= (D^\mu + a^\mu)a_\mu 
 - \sigma^2 + \omega^2 - \frac{\theta^2 }{d} - R_{\mu\nu} u^\mu u^\nu,
\label{Eq:Raychaudhuri}\\
\dot{\sigma}_{\mu\nu} &= \left(D_{\langle\mu} + a_{\langle\mu}\right)a_{\nu\rangle} 
 - \sigma_{\langle\mu}{}^\alpha\sigma_{\nu\rangle\alpha}
 + \omega_{\langle\mu}{}^\alpha\omega_{\nu\rangle\alpha}
\nonumber\\
 &-\frac{2\theta}{d}\sigma_{\mu\nu}
 - \left( R_{\mu\alpha\nu\beta} - \frac{1}{d}\Delta_{\mu\nu} R_{\alpha\beta} \right) u^\alpha u^\beta,
\label{Eq:ShearProp}\\
\dot{\omega}_{\mu\nu} &= D_{[\mu} a_{\nu]} + 2\sigma_{[\mu}{}^\alpha\omega_{\nu]\alpha} - \frac{2\theta}{d}\omega_{\mu\nu},
\label{Eq:VorticityProp}
\end{align}
where $\theta := \Delta^{\mu\nu} k_{\mu\nu} = \nabla_\mu u^\mu$, $\sigma_{\mu\nu} := k_{\langle\mu\nu\rangle}$ and $\omega_{\mu\nu} := k_{[\mu\nu]}$ denote the expansion, shear and vorticity associated with the congruence generated by $u^\mu$ and where we have abbreviated $\sigma^2 := \sigma^{\mu\nu}\sigma_{\mu\nu}$ and $\omega^2 := \omega^{\mu\nu}\omega_{\mu\nu}$. Equation~(\ref{Eq:Raychaudhuri}) is known as the Raychaudhuri equation. In the hypersurface-orthogonal case, the second term on the right-hand side of Eq.~(\ref{Eq:Curlk}) vanishes and one obtains the Codazzi–Mainardi equation. Contracting, this also yields
\begin{equation}
\Delta_\mu{}^{\mu'} G_{\mu'\nu} u^\nu = D^\nu k_{\mu\nu} - D_\mu k,
\end{equation}
which is again familiar from the $3+1$ decomposition in general relativity.

\section{Strongly hyperbolic PDE formulation}
\label{App:StronglyHyperbolicPDE}

At the beginning of Sec.~\ref{Sec:Hypo} we explained that one cannot view the derivative operators $D_\mu$ in Eq.~(\ref{Eq:FOS}) as spatial derivatives since $u^\mu$ fails to be hypersurface orthogonal in general. In this appendix we discuss how to rewrite the first-order system~(\ref{Eq:FOS}) as a strongly hyperbolic PDE system for which a well-posed Cauchy problem can be formulated based on the use of pseudodifferential operators (see Ref.~\cite{Taylor96c} and references therein).

\subsection{Strongly hyperbolic first-order systems}

We first recall some important definitions and results from Ref.~\cite{oR04} which are independent of a given spacetime foliation. Consider a first-order system of the form
\begin{equation}
\overline{\mathcal{A}}^\mu\nabla_\mu U = \overline{\mathcal{F}},
\label{Eq:FOSBis}
\end{equation}
where the components of the $m\times m$ matrix $\overline{\mathcal{A}}^\mu$ and the $m$-component vector $\overline{\mathcal{F}}$ depend smoothly on $U$ (for notational simplicity we drop the argument $U$). The associated principal symbol is defined as
\begin{equation}
\overline{\mathcal{A}}(\xi) := \overline{\mathcal{A}}^\mu\xi_\mu,
\end{equation}
for any covector field $\xi_\mu$.

\begin{definition}[\cite{oR04}]
The system~(\ref{Eq:FOSBis}) is called strongly hyperbolic if there exists a covector field $n_\mu$ such that
\begin{enumerate}
\item[(i)] $\overline{\mathcal{A}}(n)$ is invertible,
\item[(ii)] for any covector $v_\mu$ transverse to $n_\mu$, the following condition holds:
\begin{equation}
\dim\underset{\lambda\in\Real}{\oplus}\ker\left( \overline{\mathcal{A}}(\xi(\lambda)) \right) = m,\quad
\xi_\mu(\lambda) := -\lambda n_\mu + v_\mu.
\label{Eq:HypoDef}
\end{equation}
\end{enumerate}
\end{definition}
Note that $0 = \overline{\mathcal{A}}(\xi(\lambda)) U = -\lambda\overline{\mathcal{A}}(n) U + \overline{\mathcal{A}}(v) U$ is equivalent to $\lambda U = \overline{\mathcal{A}}(n)^{-1}\overline{\mathcal{A}}(v) U$, and hence $\ker\left( \overline{\mathcal{A}}(\xi(\lambda)) \right)$ is nontrivial if and only if $\lambda$ is an eigenvalue of the matrix
\begin{equation}
\hat{\mathcal{A}}(v) := \overline{\mathcal{A}}(n)^{-1}\overline{\mathcal{A}}(v).
\label{Eq:PrincipalSymbolPDE}
\end{equation}
Consequently, there are only a finite number of nontrivial terms in the sum $\oplus$ over $\lambda\in \Real$ in Eq.~(\ref{Eq:HypoDef}). Moreover, it follows that condition (ii) is equivalent to the statement that for each $v_\mu$ transverse to $n_\mu$ the matrix $\hat{\mathcal{A}}(v)$ is diagonalizable and has only real eigenvalues.

To make contact with the usual definition of strong hyperbolicity found in PDE textbooks~\cite{Kreiss89,Taylor96c}, suppose we have a foliation of spacetime by $t=const$ hypersurfaces. Let $n_\mu$ be conormal to this surfaces, such that $n_\mu$ is proportional to $\nabla_\mu t$, and let $T^\mu$ be a ``time evolution vector field" such that $T^\mu n_\mu = -1$. Finally, denote by $h_\mu{}^\nu := \delta_\mu{}^\nu + n_\mu T^\nu$ the projection operator along $n_\nu$ onto the subspace of covectors that are annihilated by $T^\mu$ (i.e. $h_\mu{}^\nu n_\nu = 0$ and $T^\mu h_\mu{}^\nu = 0$). Then, if Eq.~(\ref{Eq:FOSBis}) is strongly hyperbolic with respect to $n_\mu$, we can rewrite it as
\begin{equation}
T^\mu\nabla_\mu U = \overline{\mathcal{A}}(n)^{-1}\left[ \overline{\mathcal{A}}^\mu h_\mu{}^\nu\nabla_\nu U - \overline{\mathcal{F}} \right].
\label{Eq:FOSPDE}
\end{equation}
Since $h_\mu{}^\nu\nabla_\nu t$ is proportional to $h_\mu{}^\nu n_\nu$ which vanishes, the derivatives on the right-hand side of Eq.~(\ref{Eq:FOSPDE}) are tangent to the $t=const$ surfaces. With respect to local coordinates $x^1,x^2,\ldots,x^d$ on the $t=const$ hypersurface which are transported along $T^\mu$ such that $T^\mu\partial_\mu = \partial/\partial t$ one thus finds a PDE system of the form
\begin{equation}
\frac{\partial U}{\partial t} = \hat{\mathcal{A}}^j\frac{\partial U}{\partial x^j} + \hat{\mathcal{F}}(U),
\label{Eq:FOSPDECoords}
\end{equation}
with $\hat{\mathcal{A}}^j := \overline{\mathcal{A}}(n)^{-1} \overline{\mathcal{A}}^\mu h_\mu{}^j$ whose principal symbol is given precisely by Eq.~(\ref{Eq:PrincipalSymbolPDE}) with $v_\mu$ such that $T^\mu v_\mu = 0$. Therefore, if the system is strongly hyperbolic with respect to $n_\mu$, the principal symbol $\hat{\mathcal{A}}(v)$ is diagonalizable and has purely real eigenvalues given by $\lambda = T^\mu\xi_\mu$ with $\xi_\mu$ such that $\det(\overline{\mathcal{A}}(\xi)) = 0$.

Propositions~1 and 2 in Ref.~\cite{oR04} imply that if the system~(\ref{Eq:FOSBis}) is strongly hyperbolic with respect to some covector $n_\mu$, then it is also strongly hyperbolic with respect to covectors lying in an open set of covectors containing $n_\mu$. The boundary of this set consists of covectors $m_\mu$ such that $\overline{\mathcal{A}}(m)$ is no longer invertible. Therefore, one has a lot of freedom to choose the foliations by $t = const$ hypersurfaces; one just needs to ensure that $\overline{\mathcal{A}}(n)$ is invertible.

Based on the PDE form~(\ref{Eq:FOSPDECoords}) one can prove (local) well posedness of the Cauchy problem in the Sobolev space $H^s$ with $s > d/2 + 1$, provided the principal symbol $\hat{\mathcal{A}}(v)$ possesses a smooth symmetrizer, see, for instance, Theorem~2.3 in chapter 16 of Ref.~\cite{Taylor96c}. In particular, a smooth symmetrizer can be constructed if the eigenvalues of $\hat{\mathcal{A}}(v)$ have constant multiplicity. In this case, the desired symmetrizer is
\begin{equation}
H(v) := \sum\limits_j P_j(v)^T P_j(v),
\label{Eq:SymmetrizerPDE}
\end{equation}
where $P_j(v)$ denote the eigenprojectors associated with $\hat{\mathcal{A}}(v)$.

\subsection{Strongly hyperbolic PDEs for the fluid system}

After the remarks made in the previous subsection we are ready to apply the theory to our system~(\ref{Eq:FOS}) for the fluid equations, for which $m = (d+1)(d+3)$. In order to do so we define $\overline{\mathcal{A}}^\mu := u^\mu I - \mathcal{A}^\mu(U)$ with $I$ denoting the identity matrix and the matrices $\mathcal{A}^\mu(U)$ satisfying $\mathcal{A}^\mu(U) u_\mu = 0$. Then, Eq.~(\ref{Eq:FOS}) has precisely the form~(\ref{Eq:FOSBis}) and the associated principal symbol reads
\begin{equation}
\overline{\mathcal{A}}(\xi) 
 = \lambda I - \mathcal{A}(k),
\end{equation}
where $\xi_\mu = -\lambda u_\mu + k_\mu$ and $\mathcal{A}(k)$ is the principal symbol defined in Eq.~(\ref{Eq:PrincipalSymbol}). By choosing $n_\mu = u_\mu$ it follows that our system is strongly hyperbolic if the hypotheses of Theorem~1 are satisfied. It follows from the previous arguments that the corresponding PDE problem is well posed as long as the foliation is chosen such that the matrix $\overline{\mathcal{A}}(n)$ associated with the normal covector $n_\mu$ is invertible and a smooth symmetrizer for $\hat{\mathcal{A}}(v)$ can be constructed.

The eigenvalues $\Lambda = T^\mu\xi_\mu$ of $\hat{\mathcal{A}}(v)$ are related to those of $\mathcal{A}(k)$ which have been computed and denoted by $\lambda$ in Sec.~\ref{Sec:Hypo}, according to
\begin{equation}
\Lambda = \gamma^{-1}\lambda + T^\mu k_\mu,\qquad
\gamma^{-1}:=-T^\mu u_\mu \in (0,1].
\end{equation}
It follows that two eigenvalues $\lambda$ and $\lambda'$ of $\mathcal{A}(k)$ are distinct from each other if and only if the corresponding eigenvalues $\Lambda$ and $\Lambda'$ of $\hat{\mathcal{A}}(v)$ are different from each other. Therefore, assuming that the eigenvalues $\lambda$ of $\mathcal{A}(k)$ (which depend only on the temperature) do not cross, the same is true for the eigenvalues $\Lambda$ of the PDE symbol $\hat{\mathcal{A}}(v)$ and one can construct a smooth symmetrizer as in Eq.~(\ref{Eq:SymmetrizerPDE}).

\subsection{Causality}

Finally, we remark that the covectors $\xi_\mu$ for which $\overline{\mathcal{A}}(\xi)$ has a nontrivial kernel are of the form
\begin{equation}
\xi_\mu = -\lambda u_\mu + k_\mu,\qquad u^\mu k_\mu = 0,\quad k^\mu k_\mu = 1,
\end{equation}
with $\lambda$ the eigenvalues computed in Sec.~\ref{Sec:Hypo}. Since the conditions for causality imply that $|\lambda|\leq 1$, it follows that $\xi^\mu\xi_\mu = 1 -\lambda^2 \geq 0$, and hence $\xi^\mu$ must be null or spacelike. Consequently, it follows that $\overline{\mathcal{A}}(n)$ cannot have nontrivial kernel if $n$ is \emph{timelike}, and hence $\overline{\mathcal{A}}(n)$ is invertible for all timelike unit vectors $n$. This implies that one can pose the PDE problem~(\ref{Eq:FOSPDECoords}) on any foliation by spacelike hypersurfaces $t=const$ on the spacetime manifold $(M,g)$.

\bibliography{refs_kinetic}

\begin{thebibliography}{44}%
\makeatletter
\providecommand \@ifxundefined [1]{%
 \@ifx{#1\undefined}
}%
\providecommand \@ifnum [1]{%
 \ifnum #1\expandafter \@firstoftwo
 \else \expandafter \@secondoftwo
 \fi
}%
\providecommand \@ifx [1]{%
 \ifx #1\expandafter \@firstoftwo
 \else \expandafter \@secondoftwo
 \fi
}%
\providecommand \natexlab [1]{#1}%
\providecommand \enquote  [1]{``#1''}%
\providecommand \bibnamefont  [1]{#1}%
\providecommand \bibfnamefont [1]{#1}%
\providecommand \citenamefont [1]{#1}%
\providecommand \href@noop [0]{\@secondoftwo}%
\providecommand \href [0]{\begingroup \@sanitize@url \@href}%
\providecommand \@href[1]{\@@startlink{#1}\@@href}%
\providecommand \@@href[1]{\endgroup#1\@@endlink}%
\providecommand \@sanitize@url [0]{\catcode `\\12\catcode `\$12\catcode
  `\&12\catcode `\#12\catcode `\^12\catcode `\_12\catcode `\%12\relax}%
\providecommand \@@startlink[1]{}%
\providecommand \@@endlink[0]{}%
\providecommand \url  [0]{\begingroup\@sanitize@url \@url }%
\providecommand \@url [1]{\endgroup\@href {#1}{\urlprefix }}%
\providecommand \urlprefix  [0]{URL }%
\providecommand \Eprint [0]{\href }%
\providecommand \doibase [0]{https://doi.org/}%
\providecommand \selectlanguage [0]{\@gobble}%
\providecommand \bibinfo  [0]{\@secondoftwo}%
\providecommand \bibfield  [0]{\@secondoftwo}%
\providecommand \translation [1]{[#1]}%
\providecommand \BibitemOpen [0]{}%
\providecommand \bibitemStop [0]{}%
\providecommand \bibitemNoStop [0]{.\EOS\space}%
\providecommand \EOS [0]{\spacefactor3000\relax}%
\providecommand \BibitemShut  [1]{\csname bibitem#1\endcsname}%
\let\auto@bib@innerbib\@empty
\bibitem [{\citenamefont {{Derradi de Souza}}\ \emph
  {et~al.}(2016)\citenamefont {{Derradi de Souza}}, \citenamefont {Koide},\
  and\ \citenamefont {Kodama}}]{DKK2016}%
  \BibitemOpen
  \bibfield  {author} {\bibinfo {author} {\bibfnamefont {R.}~\bibnamefont
  {{Derradi de Souza}}}, \bibinfo {author} {\bibfnamefont {T.}~\bibnamefont
  {Koide}},\ and\ \bibinfo {author} {\bibfnamefont {T.}~\bibnamefont
  {Kodama}},\ }\bibfield  {title} {\bibinfo {title} {Hydrodynamic approaches in
  relativistic heavy ion reactions},\ }\href
  {https://doi.org/https://doi.org/10.1016/j.ppnp.2015.09.002} {\bibfield
  {journal} {\bibinfo  {journal} {Progress in Particle and Nuclear Physics}\
  }\textbf {\bibinfo {volume} {86}},\ \bibinfo {pages} {35} (\bibinfo {year}
  {2016})}\BibitemShut {NoStop}%
\bibitem [{\citenamefont {Du}\ and\ \citenamefont {Heinz}(2020)}]{DH2020}%
  \BibitemOpen
  \bibfield  {author} {\bibinfo {author} {\bibfnamefont {L.}~\bibnamefont
  {Du}}\ and\ \bibinfo {author} {\bibfnamefont {U.}~\bibnamefont {Heinz}},\
  }\bibfield  {title} {\bibinfo {title} {(3+1)-dimensional dissipative
  relativistic fluid dynamics at non-zero net baryon density},\ }\href
  {https://doi.org/https://doi.org/10.1016/j.cpc.2019.107090} {\bibfield
  {journal} {\bibinfo  {journal} {Computer Physics Communications}\ }\textbf
  {\bibinfo {volume} {251}},\ \bibinfo {pages} {107090} (\bibinfo {year}
  {2020})}\BibitemShut {NoStop}%
\bibitem [{\citenamefont {Shen}\ and\ \citenamefont {Yan}(2020)}]{Shen2020}%
  \BibitemOpen
  \bibfield  {author} {\bibinfo {author} {\bibfnamefont {C.}~\bibnamefont
  {Shen}}\ and\ \bibinfo {author} {\bibfnamefont {L.}~\bibnamefont {Yan}},\
  }\bibfield  {title} {\bibinfo {title} {Recent development of hydrodynamic
  modeling in heavy-ion collisions},\ }\href
  {https://doi.org/10.1007/s41365-020-00829-z} {\bibfield  {journal} {\bibinfo
  {journal} {Nuclear Science and Techniques}\ }\textbf {\bibinfo {volume}
  {31}},\ \bibinfo {pages} {122} (\bibinfo {year} {2020})}\BibitemShut
  {NoStop}%
\bibitem [{\citenamefont {Abbott}\ \emph
  {et~al.}(2017{\natexlab{a}})\citenamefont {Abbott}, \citenamefont {Abbott},
  \citenamefont {Abbott},\ and\ \citenamefont {et~al.}}]{LIGO2017_A}%
  \BibitemOpen
  \bibfield  {author} {\bibinfo {author} {\bibfnamefont {B.~P.}\ \bibnamefont
  {Abbott}}, \bibinfo {author} {\bibfnamefont {R.}~\bibnamefont {Abbott}},
  \bibinfo {author} {\bibfnamefont {T.~D.}\ \bibnamefont {Abbott}},\ and\
  \bibinfo {author} {\bibfnamefont {F.~A.}\ \bibnamefont {et~al.}} (\bibinfo
  {collaboration} {LIGO Scientific Collaboration and Virgo Collaboration}),\
  }\bibfield  {title} {\bibinfo {title} {{GW}170817: {O}bservation of
  {G}ravitational {W}aves from a {B}inary {N}eutron {S}tar {I}nspiral},\ }\href
  {https://doi.org/10.1103/PhysRevLett.119.161101} {\bibfield  {journal}
  {\bibinfo  {journal} {Phys. Rev. Lett.}\ }\textbf {\bibinfo {volume} {119}},\
  \bibinfo {pages} {161101} (\bibinfo {year} {2017}{\natexlab{a}})}\BibitemShut
  {NoStop}%
\bibitem [{\citenamefont {Abbott}\ \emph
  {et~al.}(2017{\natexlab{b}})\citenamefont {Abbott}, \citenamefont {Abbott},
  \citenamefont {Abbott},\ and\ \citenamefont {et~al}}]{LIGO2017_B}%
  \BibitemOpen
  \bibfield  {author} {\bibinfo {author} {\bibfnamefont {B.~P.}\ \bibnamefont
  {Abbott}}, \bibinfo {author} {\bibfnamefont {R.}~\bibnamefont {Abbott}},
  \bibinfo {author} {\bibfnamefont {T.~D.}\ \bibnamefont {Abbott}},\ and\
  \bibinfo {author} {\bibfnamefont {F.~A.}\ \bibnamefont {et~al}},\ }\bibfield
  {title} {\bibinfo {title} {Gravitational {W}aves and {G}amma-{R}ays from a
  {B}inary {N}eutron {S}tar {M}erger: {GW}170817 and {GRB} 170817a},\ }\href
  {https://doi.org/10.3847/2041-8213/aa920c} {\bibfield  {journal} {\bibinfo
  {journal} {The Astrophysical Journal Letters}\ }\textbf {\bibinfo {volume}
  {848}},\ \bibinfo {pages} {L13} (\bibinfo {year}
  {2017}{\natexlab{b}})}\BibitemShut {NoStop}%
\bibitem [{\citenamefont {Abbott}\ \emph {et~al.}(2021)\citenamefont {Abbott},
  \citenamefont {Abbott}, \citenamefont {Abbott},\ and\ \citenamefont
  {et~al.}}]{LIGO2021}%
  \BibitemOpen
  \bibfield  {author} {\bibinfo {author} {\bibfnamefont {B.~P.}\ \bibnamefont
  {Abbott}}, \bibinfo {author} {\bibfnamefont {R.}~\bibnamefont {Abbott}},
  \bibinfo {author} {\bibfnamefont {T.~D.}\ \bibnamefont {Abbott}},\ and\
  \bibinfo {author} {\bibfnamefont {F.~A.}\ \bibnamefont {et~al.}} (\bibinfo
  {collaboration} {LIGO Scientific Collaboration, the Virgo Collaboration and
  the KAGRA Collaboration}),\ }\bibfield  {title} {\bibinfo {title}
  {Observation of {G}ravitational {W}aves from {T}wo {N}eutron {S}tar–{B}lack
  {H}ole {C}oalescences},\ }\href {https://doi.org/10.3847/2041-8213/ac082e}
  {\bibfield  {journal} {\bibinfo  {journal} {The Astrophysical Journal
  Letters}\ }\textbf {\bibinfo {volume} {915}},\ \bibinfo {pages} {L5}
  (\bibinfo {year} {2021})}\BibitemShut {NoStop}%
\bibitem [{\citenamefont {Chabanov}\ \emph {et~al.}(2021)\citenamefont
  {Chabanov}, \citenamefont {Rezzolla},\ and\ \citenamefont
  {Rischke}}]{ChR2021}%
  \BibitemOpen
  \bibfield  {author} {\bibinfo {author} {\bibfnamefont {M.}~\bibnamefont
  {Chabanov}}, \bibinfo {author} {\bibfnamefont {L.}~\bibnamefont {Rezzolla}},\
  and\ \bibinfo {author} {\bibfnamefont {D.~H.}\ \bibnamefont {Rischke}},\
  }\bibfield  {title} {\bibinfo {title} {{General-relativistic hydrodynamics of
  non-perfect fluids: 3+1 conservative formulation and application to viscous
  black hole accretion}},\ }\href {https://doi.org/10.1093/mnras/stab1384}
  {\bibfield  {journal} {\bibinfo  {journal} {Monthly Notices of the Royal
  Astronomical Society}\ }\textbf {\bibinfo {volume} {505}},\ \bibinfo {pages}
  {5910} (\bibinfo {year} {2021})}\BibitemShut {NoStop}%
\bibitem [{\citenamefont {Hatton}\ and\ \citenamefont
  {Hawke}(2024)}]{Hatton2024}%
  \BibitemOpen
  \bibfield  {author} {\bibinfo {author} {\bibfnamefont {M.~J.}\ \bibnamefont
  {Hatton}}\ and\ \bibinfo {author} {\bibfnamefont {I.}~\bibnamefont {Hawke}},\
  }\bibfield  {title} {\bibinfo {title} {{A dissipative extension to ideal
  hydrodynamics}},\ }\href {https://doi.org/10.1093/mnras/stae2284} {\bibfield
  {journal} {\bibinfo  {journal} {Monthly Notices of the Royal Astronomical
  Society}\ }\textbf {\bibinfo {volume} {535}},\ \bibinfo {pages} {47}
  (\bibinfo {year} {2024})}\BibitemShut {NoStop}%
\bibitem [{\citenamefont {Ván}(2020)}]{Van2020}%
  \BibitemOpen
  \bibfield  {author} {\bibinfo {author} {\bibfnamefont {P.}~\bibnamefont
  {Ván}},\ }\bibfield  {title} {\bibinfo {title} {Nonequilibrium
  thermodynamics: emergent and fundamental},\ }\href
  {https://doi.org/10.1098/rsta.2020.0066} {\bibfield  {journal} {\bibinfo
  {journal} {Philosophical Transactions of the Royal Society A: Mathematical,
  Physical and Engineering Sciences}\ }\textbf {\bibinfo {volume} {378}},\
  \bibinfo {pages} {20200066} (\bibinfo {year} {2020})}\BibitemShut {NoStop}%
\bibitem [{\citenamefont {Salazar}\ and\ \citenamefont
  {Zannias}(2020)}]{SZ2020}%
  \BibitemOpen
  \bibfield  {author} {\bibinfo {author} {\bibfnamefont {J.~F.}\ \bibnamefont
  {Salazar}}\ and\ \bibinfo {author} {\bibfnamefont {T.}~\bibnamefont
  {Zannias}},\ }\bibfield  {title} {\bibinfo {title} {On extended
  thermodynamics: From classical to the relativistic regime},\ }\href
  {https://doi.org/10.1142/S0218271820300104} {\bibfield  {journal} {\bibinfo
  {journal} {International Journal of Modern Physics D}\ }\textbf {\bibinfo
  {volume} {29}},\ \bibinfo {pages} {2030010} (\bibinfo {year}
  {2020})}\BibitemShut {NoStop}%
\bibitem [{\citenamefont {Gavassino}\ and\ \citenamefont
  {Antonelli}(2021)}]{Gavassino2021}%
  \BibitemOpen
  \bibfield  {author} {\bibinfo {author} {\bibfnamefont {L.}~\bibnamefont
  {Gavassino}}\ and\ \bibinfo {author} {\bibfnamefont {M.}~\bibnamefont
  {Antonelli}},\ }\bibfield  {title} {\bibinfo {title} {Unified extended
  irreversible thermodynamics and the stability of relativistic theories for
  dissipation},\ }\bibfield  {journal} {\bibinfo  {journal} {Frontiers in
  Astronomy and Space Sciences}\ }\textbf {\bibinfo {volume} {8}},\ \href
  {https://doi.org/10.3389/fspas.2021.686344} {10.3389/fspas.2021.686344}
  (\bibinfo {year} {2021})\BibitemShut {NoStop}%
\bibitem [{\citenamefont {Rocha}\ \emph {et~al.}(2024)\citenamefont {Rocha},
  \citenamefont {Wagner}, \citenamefont {Denicol}, \citenamefont {Noronha},\
  and\ \citenamefont {Rischke}}]{RDNR2024}%
  \BibitemOpen
  \bibfield  {author} {\bibinfo {author} {\bibfnamefont {G.~S.}\ \bibnamefont
  {Rocha}}, \bibinfo {author} {\bibfnamefont {D.}~\bibnamefont {Wagner}},
  \bibinfo {author} {\bibfnamefont {G.~S.}\ \bibnamefont {Denicol}}, \bibinfo
  {author} {\bibfnamefont {J.}~\bibnamefont {Noronha}},\ and\ \bibinfo {author}
  {\bibfnamefont {D.~H.}\ \bibnamefont {Rischke}},\ }\bibfield  {title}
  {\bibinfo {title} {Theories of relativistic dissipative fluid dynamics},\
  }\bibfield  {journal} {\bibinfo  {journal} {Entropy}\ }\textbf {\bibinfo
  {volume} {26}},\ \href {https://doi.org/10.3390/e26030189}
  {10.3390/e26030189} (\bibinfo {year} {2024})\BibitemShut {NoStop}%
\bibitem [{\citenamefont {Bemfica}\ \emph {et~al.}(2018)\citenamefont
  {Bemfica}, \citenamefont {Disconzi},\ and\ \citenamefont
  {Noronha}}]{BDN2018}%
  \BibitemOpen
  \bibfield  {author} {\bibinfo {author} {\bibfnamefont {F.~S.}\ \bibnamefont
  {Bemfica}}, \bibinfo {author} {\bibfnamefont {M.~M.}\ \bibnamefont
  {Disconzi}},\ and\ \bibinfo {author} {\bibfnamefont {J.}~\bibnamefont
  {Noronha}},\ }\bibfield  {title} {\bibinfo {title} {Causality and existence
  of solutions of relativistic viscous fluid dynamics with gravity},\ }\href
  {https://doi.org/10.1103/PhysRevD.98.104064} {\bibfield  {journal} {\bibinfo
  {journal} {Phys. Rev. D}\ }\textbf {\bibinfo {volume} {98}},\ \bibinfo
  {pages} {104064} (\bibinfo {year} {2018})}\BibitemShut {NoStop}%
\bibitem [{\citenamefont {Bemfica}\ \emph {et~al.}(2019)\citenamefont
  {Bemfica}, \citenamefont {Disconzi},\ and\ \citenamefont
  {Noronha}}]{BDN2019}%
  \BibitemOpen
  \bibfield  {author} {\bibinfo {author} {\bibfnamefont {F.~S.}\ \bibnamefont
  {Bemfica}}, \bibinfo {author} {\bibfnamefont {M.~M.}\ \bibnamefont
  {Disconzi}},\ and\ \bibinfo {author} {\bibfnamefont {J.}~\bibnamefont
  {Noronha}},\ }\bibfield  {title} {\bibinfo {title} {Nonlinear causality of
  general first-order relativistic viscous hydrodynamics},\ }\href
  {https://doi.org/10.1103/PhysRevD.100.104020} {\bibfield  {journal} {\bibinfo
   {journal} {Phys. Rev. D}\ }\textbf {\bibinfo {volume} {100}},\ \bibinfo
  {pages} {104020} (\bibinfo {year} {2019})}\BibitemShut {NoStop}%
\bibitem [{\citenamefont {Kovtun}(2019)}]{pK19}%
  \BibitemOpen
  \bibfield  {author} {\bibinfo {author} {\bibfnamefont {P.}~\bibnamefont
  {Kovtun}},\ }\bibfield  {title} {\bibinfo {title} {First-order relativistic
  hydrodynamics is stable},\ }\href {https://doi.org/10.1007/JHEP10(2019)034}
  {\bibfield  {journal} {\bibinfo  {journal} {JHEP}\ }\textbf {\bibinfo
  {volume} {10}},\ \bibinfo {pages} {034}}\BibitemShut {NoStop}%
\bibitem [{\citenamefont {Hoult}\ and\ \citenamefont
  {Kovtun}(2020)}]{Hoult2020}%
  \BibitemOpen
  \bibfield  {author} {\bibinfo {author} {\bibfnamefont {R.~E.}\ \bibnamefont
  {Hoult}}\ and\ \bibinfo {author} {\bibfnamefont {P.}~\bibnamefont {Kovtun}},\
  }\bibfield  {title} {\bibinfo {title} {Stable and causal relativistic
  {N}avier-{S}tokes equations},\ }\href
  {https://doi.org/10.1007/JHEP06(2020)067} {\bibfield  {journal} {\bibinfo
  {journal} {Journal of High Energy Physics}\ }\textbf {\bibinfo {volume}
  {2020}},\ \bibinfo {pages} {67} (\bibinfo {year} {2020})}\BibitemShut
  {NoStop}%
\bibitem [{\citenamefont {Bemfica}\ \emph {et~al.}(2022)\citenamefont
  {Bemfica}, \citenamefont {Disconzi},\ and\ \citenamefont
  {Noronha}}]{Bemfica2022}%
  \BibitemOpen
  \bibfield  {author} {\bibinfo {author} {\bibfnamefont {F.~S.}\ \bibnamefont
  {Bemfica}}, \bibinfo {author} {\bibfnamefont {M.~M.}\ \bibnamefont
  {Disconzi}},\ and\ \bibinfo {author} {\bibfnamefont {J.}~\bibnamefont
  {Noronha}},\ }\bibfield  {title} {\bibinfo {title} {First-order
  general-relativistic viscous fluid dynamics},\ }\href
  {https://doi.org/10.1103/PhysRevX.12.021044} {\bibfield  {journal} {\bibinfo
  {journal} {Phys. Rev. X}\ }\textbf {\bibinfo {volume} {12}},\ \bibinfo
  {pages} {021044} (\bibinfo {year} {2022})}\BibitemShut {NoStop}%
\bibitem [{\citenamefont {Hoult}\ and\ \citenamefont
  {Kovtun}(2022)}]{Kovtun2022}%
  \BibitemOpen
  \bibfield  {author} {\bibinfo {author} {\bibfnamefont {R.~E.}\ \bibnamefont
  {Hoult}}\ and\ \bibinfo {author} {\bibfnamefont {P.}~\bibnamefont {Kovtun}},\
  }\bibfield  {title} {\bibinfo {title} {Causal first-order hydrodynamics from
  kinetic theory and holography},\ }\href
  {https://doi.org/10.1103/PhysRevD.106.066023} {\bibfield  {journal} {\bibinfo
   {journal} {Phys. Rev. D}\ }\textbf {\bibinfo {volume} {106}},\ \bibinfo
  {pages} {066023} (\bibinfo {year} {2022})}\BibitemShut {NoStop}%
\bibitem [{\citenamefont {Rocha}\ \emph {et~al.}(2022)\citenamefont {Rocha},
  \citenamefont {Denicol},\ and\ \citenamefont {Noronha}}]{Rocha2022}%
  \BibitemOpen
  \bibfield  {author} {\bibinfo {author} {\bibfnamefont {G.~S.}\ \bibnamefont
  {Rocha}}, \bibinfo {author} {\bibfnamefont {G.~S.}\ \bibnamefont {Denicol}},\
  and\ \bibinfo {author} {\bibfnamefont {J.}~\bibnamefont {Noronha}},\
  }\bibfield  {title} {\bibinfo {title} {Perturbative approaches in
  relativistic kinetic theory and the emergence of first-order hydrodynamics},\
  }\href {https://doi.org/10.1103/PhysRevD.106.036010} {\bibfield  {journal}
  {\bibinfo  {journal} {Phys. Rev. D}\ }\textbf {\bibinfo {volume} {106}},\
  \bibinfo {pages} {036010} (\bibinfo {year} {2022})}\BibitemShut {NoStop}%
\bibitem [{\citenamefont {Disconzi}(2024)}]{Disconzi2024}%
  \BibitemOpen
  \bibfield  {author} {\bibinfo {author} {\bibfnamefont {M.}~\bibnamefont
  {Disconzi}},\ }\bibfield  {title} {\bibinfo {title} {Recent developments in
  mathematical aspects of relativistic fluids},\ }\href@noop {} {\bibfield
  {journal} {\bibinfo  {journal} {Living Reviews in Relativity}\ }\textbf
  {\bibinfo {volume} {27}},\ \bibinfo {pages} {6} (\bibinfo {year}
  {2024})}\BibitemShut {NoStop}%
\bibitem [{\citenamefont {Hiscock}\ and\ \citenamefont
  {Lindblom}(1985)}]{whlL85}%
  \BibitemOpen
  \bibfield  {author} {\bibinfo {author} {\bibfnamefont {W.}~\bibnamefont
  {Hiscock}}\ and\ \bibinfo {author} {\bibfnamefont {L.}~\bibnamefont
  {Lindblom}},\ }\bibfield  {title} {\bibinfo {title} {Generic instabilities in
  first-order dissipative relativistic fluid theories},\ }\href
  {https://doi.org/10.1103/PhysRevD.31.725} {\bibfield  {journal} {\bibinfo
  {journal} {Phys. Rev. D}\ }\textbf {\bibinfo {volume} {31}},\ \bibinfo
  {pages} {725} (\bibinfo {year} {1985})}\BibitemShut {NoStop}%
\bibitem [{\citenamefont {Ciambelli}\ and\ \citenamefont
  {Lehner}(2023)}]{lClL23}%
  \BibitemOpen
  \bibfield  {author} {\bibinfo {author} {\bibfnamefont {L.}~\bibnamefont
  {Ciambelli}}\ and\ \bibinfo {author} {\bibfnamefont {L.}~\bibnamefont
  {Lehner}},\ }\bibfield  {title} {\bibinfo {title} {{Fluid-gravity
  correspondence and causal first-order relativistic viscous hydrodynamics}},\
  }\href {https://doi.org/10.1103/PhysRevD.108.126019} {\bibfield  {journal}
  {\bibinfo  {journal} {Phys. Rev. D}\ }\textbf {\bibinfo {volume} {108}},\
  \bibinfo {pages} {126019} (\bibinfo {year} {2023})}\BibitemShut {NoStop}%
\bibitem [{\citenamefont {Stewart}(1971)}]{Stewart1971}%
  \BibitemOpen
  \bibfield  {author} {\bibinfo {author} {\bibfnamefont {J.~M.}\ \bibnamefont
  {Stewart}},\ }\bibinfo {title} {Non-equilibrium relativistic kinetic
  theory},\ in\ \href {https://doi.org/10.1007/BFb0025375} {\emph {\bibinfo
  {booktitle} {Non-Equilibrium Relativistic Kinetic Theory}}}\ (\bibinfo
  {publisher} {Springer Berlin Heidelberg},\ \bibinfo {address} {Berlin,
  Heidelberg},\ \bibinfo {year} {1971})\ pp.\ \bibinfo {pages}
  {1--113}\BibitemShut {NoStop}%
\bibitem [{\citenamefont {Salazar}\ \emph {et~al.}(2025)\citenamefont
  {Salazar}, \citenamefont {Garc\'ia-Perciante},\ and\ \citenamefont
  {Sarbach}}]{aGjSoS2024a}%
  \BibitemOpen
  \bibfield  {author} {\bibinfo {author} {\bibfnamefont {J.~F.}\ \bibnamefont
  {Salazar}}, \bibinfo {author} {\bibfnamefont {A.~L.}\ \bibnamefont
  {Garc\'ia-Perciante}},\ and\ \bibinfo {author} {\bibfnamefont
  {O.}~\bibnamefont {Sarbach}},\ }\bibfield  {title} {\bibinfo {title}
  {Relativistic dissipative fluids in the trace-fixed particle frame:
  Hyperbolicity, causality, and stability},\ }\href
  {https://doi.org/10.1103/PhysRevD.111.L081501} {\bibfield  {journal}
  {\bibinfo  {journal} {Phys. Rev. D}\ }\textbf {\bibinfo {volume} {111}},\
  \bibinfo {pages} {L081501} (\bibinfo {year} {2025})}\BibitemShut {NoStop}%
\bibitem [{\citenamefont {Gabarrete}\ \emph {et~al.}()\citenamefont
  {Gabarrete}, \citenamefont {Garc\'ia-Perciante},\ and\ \citenamefont
  {Sarbach}}]{cGaGoS25}%
  \BibitemOpen
  \bibfield  {author} {\bibinfo {author} {\bibfnamefont {C.}~\bibnamefont
  {Gabarrete}}, \bibinfo {author} {\bibfnamefont {A.~L.}\ \bibnamefont
  {Garc\'ia-Perciante}},\ and\ \bibinfo {author} {\bibfnamefont
  {O.}~\bibnamefont {Sarbach}},\ }\href@noop {} {\ }\bibinfo {note} {In
  preparation}\BibitemShut {NoStop}%
\bibitem [{\citenamefont {Kreiss}\ and\ \citenamefont
  {Lorenz}(1989)}]{Kreiss89}%
  \BibitemOpen
  \bibfield  {author} {\bibinfo {author} {\bibfnamefont {H.~O.}\ \bibnamefont
  {Kreiss}}\ and\ \bibinfo {author} {\bibfnamefont {J.}~\bibnamefont
  {Lorenz}},\ }\href@noop {} {\emph {\bibinfo {title} {Initial-boundary value
  problems and the {N}avier-{S}tokes equations}}}\ (\bibinfo  {publisher}
  {Academic Press},\ \bibinfo {address} {San Diego},\ \bibinfo {year}
  {1989})\BibitemShut {NoStop}%
\bibitem [{\citenamefont {Nagy}\ \emph {et~al.}(2004)\citenamefont {Nagy},
  \citenamefont {Ortiz},\ and\ \citenamefont {Reula}}]{gNoOoR04}%
  \BibitemOpen
  \bibfield  {author} {\bibinfo {author} {\bibfnamefont {G.}~\bibnamefont
  {Nagy}}, \bibinfo {author} {\bibfnamefont {O.}~\bibnamefont {Ortiz}},\ and\
  \bibinfo {author} {\bibfnamefont {O.}~\bibnamefont {Reula}},\ }\bibfield
  {title} {\bibinfo {title} {Strongly hyperbolic second order {E}instein's
  evolution equations},\ }\href@noop {} {\bibfield  {journal} {\bibinfo
  {journal} {Phys. Rev.}\ }\textbf {\bibinfo {volume} {D70}},\ \bibinfo {pages}
  {044012} (\bibinfo {year} {2004})}\BibitemShut {NoStop}%
\bibitem [{\citenamefont {Sarbach}\ and\ \citenamefont
  {Tiglio}(2012)}]{oSmT12}%
  \BibitemOpen
  \bibfield  {author} {\bibinfo {author} {\bibfnamefont {O.}~\bibnamefont
  {Sarbach}}\ and\ \bibinfo {author} {\bibfnamefont {M.}~\bibnamefont
  {Tiglio}},\ }\bibfield  {title} {\bibinfo {title} {{Continuum and Discrete
  Initial-Boundary-Value Problems and Einstein's Field Equations}},\ }\href
  {https://doi.org/10.12942/lrr-2012-9} {\bibfield  {journal} {\bibinfo
  {journal} {Living Rev. Rel.}\ }\textbf {\bibinfo {volume} {15}},\ \bibinfo
  {pages} {9} (\bibinfo {year} {2012})}\BibitemShut {NoStop}%
\bibitem [{\citenamefont {Hawking}\ and\ \citenamefont
  {Ellis}(1973)}]{HawkingEllis-Book}%
  \BibitemOpen
  \bibfield  {author} {\bibinfo {author} {\bibfnamefont {S.}~\bibnamefont
  {Hawking}}\ and\ \bibinfo {author} {\bibfnamefont {G.}~\bibnamefont
  {Ellis}},\ }\href@noop {} {\emph {\bibinfo {title} {The Large Scale Structure
  of Space Time}}}\ (\bibinfo  {publisher} {Cambridge University Press},\
  \bibinfo {address} {Cambridge},\ \bibinfo {year} {1973})\BibitemShut
  {NoStop}%
\bibitem [{\citenamefont {Wald}(1984)}]{Wald1984}%
  \BibitemOpen
  \bibfield  {author} {\bibinfo {author} {\bibfnamefont {R.~M.}\ \bibnamefont
  {Wald}},\ }\href {https://doi.org/10.7208/chicago/9780226870373.001.0001}
  {\emph {\bibinfo {title} {{General Relativity}}}}\ (\bibinfo  {publisher}
  {Chicago Univ. Pr.},\ \bibinfo {address} {Chicago, USA},\ \bibinfo {year}
  {1984})\BibitemShut {NoStop}%
\bibitem [{\citenamefont {Ellis}\ \emph {et~al.}(2012)\citenamefont {Ellis},
  \citenamefont {Maartens},\ and\ \citenamefont {MacCallum}}]{Ellis2012}%
  \BibitemOpen
  \bibfield  {author} {\bibinfo {author} {\bibfnamefont {G.~F.~R.}\
  \bibnamefont {Ellis}}, \bibinfo {author} {\bibfnamefont {R.}~\bibnamefont
  {Maartens}},\ and\ \bibinfo {author} {\bibfnamefont {M.~A.~H.}\ \bibnamefont
  {MacCallum}},\ }\href@noop {} {\emph {\bibinfo {title} {Relativistic
  Cosmology}}}\ (\bibinfo  {publisher} {Cambridge University Press},\ \bibinfo
  {year} {2012})\BibitemShut {NoStop}%
\bibitem [{\citenamefont {Abalos}\ and\ \citenamefont {Reula}(2020)}]{fAoR20}%
  \BibitemOpen
  \bibfield  {author} {\bibinfo {author} {\bibfnamefont {F.}~\bibnamefont
  {Abalos}}\ and\ \bibinfo {author} {\bibfnamefont {O.}~\bibnamefont {Reula}},\
  }\bibfield  {title} {\bibinfo {title} {On necessary and sufficient conditions
  for strong hyperbolicity in systems with constraints},\ }\href
  {https://doi.org/10.1088/1361-6382/ab954c} {\bibfield  {journal} {\bibinfo
  {journal} {Class. Quant. Grav.}\ }\textbf {\bibinfo {volume} {37}},\ \bibinfo
  {pages} {185012} (\bibinfo {year} {2020})}\BibitemShut {NoStop}%
\bibitem [{\citenamefont {Abalos}(2022)}]{fA22}%
  \BibitemOpen
  \bibfield  {author} {\bibinfo {author} {\bibfnamefont {J.}~\bibnamefont
  {Abalos}},\ }\bibfield  {title} {\bibinfo {title} {On constraint preservation
  and strong hyperbolicity},\ }\href {https://doi.org/10.1088/1361-6382/ac88af}
  {\bibfield  {journal} {\bibinfo  {journal} {Class. Quant. Grav.}\ }\textbf
  {\bibinfo {volume} {39}},\ \bibinfo {pages} {215004} (\bibinfo {year}
  {2022})}\BibitemShut {NoStop}%
\bibitem [{\citenamefont {Abalos}\ \emph {et~al.}(2024)\citenamefont {Abalos},
  \citenamefont {Reula},\ and\ \citenamefont {Hilditch}}]{fAoRdH24}%
  \BibitemOpen
  \bibfield  {author} {\bibinfo {author} {\bibfnamefont {F.}~\bibnamefont
  {Abalos}}, \bibinfo {author} {\bibfnamefont {O.}~\bibnamefont {Reula}},\ and\
  \bibinfo {author} {\bibfnamefont {D.}~\bibnamefont {Hilditch}},\ }\bibfield
  {title} {\bibinfo {title} {Hyperbolic extensions of constrained {PDE}s},\
  }\href@noop {} {\  (\bibinfo {year} {2024})},\ \Eprint
  {https://arxiv.org/abs/2410.18286} {arXiv:2410.18286 [math.AP]} \BibitemShut
  {NoStop}%
\bibitem [{\citenamefont {Geroch}(1996)}]{rG96}%
  \BibitemOpen
  \bibfield  {author} {\bibinfo {author} {\bibfnamefont {R.}~\bibnamefont
  {Geroch}},\ }\bibfield  {title} {\bibinfo {title} {Partial differential
  equations of physics},\ }\href@noop {} {\bibfield  {journal} {\bibinfo
  {journal} {General Relativity: Proceedings. Edited by G.S. Hall and J.R.
  Pulham. Edinburgh, IOP Publishing}\ ,\ \bibinfo {pages} {19}} (\bibinfo
  {year} {1996})},\ \Eprint {https://arxiv.org/abs/gr-qc/9602055}
  {arXiv:gr-qc/9602055} \BibitemShut {NoStop}%
\bibitem [{\citenamefont {Reula}(2004)}]{oR04}%
  \BibitemOpen
  \bibfield  {author} {\bibinfo {author} {\bibfnamefont {O.}~\bibnamefont
  {Reula}},\ }\bibfield  {title} {\bibinfo {title} {{Strongly hyperbolic
  systems in general relativity}},\ }\href
  {https://doi.org/10.1142/S0219891604000111} {\bibfield  {journal} {\bibinfo
  {journal} {Diff. Eq.}\ }\textbf {\bibinfo {volume} {01}},\ \bibinfo {pages}
  {251} (\bibinfo {year} {2004})}\BibitemShut {NoStop}%
\bibitem [{\citenamefont {Eckart}(1940)}]{EckartIII}%
  \BibitemOpen
  \bibfield  {author} {\bibinfo {author} {\bibfnamefont {C.}~\bibnamefont
  {Eckart}},\ }\bibfield  {title} {\bibinfo {title} {The thermodynamics of
  irreversible processes. iii. relativistic theory of the simple fluid},\
  }\href {https://doi.org/10.1103/PhysRev.58.919} {\bibfield  {journal}
  {\bibinfo  {journal} {Phys. Rev.}\ }\textbf {\bibinfo {volume} {58}},\
  \bibinfo {pages} {919} (\bibinfo {year} {1940})}\BibitemShut {NoStop}%
\bibitem [{\citenamefont {Garc\'ia-Perciante}\ \emph
  {et~al.}(2025)\citenamefont {Garc\'ia-Perciante}, \citenamefont {M\'endez},\
  and\ \citenamefont {Sarbach}}]{JNET24}%
  \BibitemOpen
  \bibfield  {author} {\bibinfo {author} {\bibfnamefont {A.~L.}\ \bibnamefont
  {Garc\'ia-Perciante}}, \bibinfo {author} {\bibfnamefont {A.~R.}\ \bibnamefont
  {M\'endez}},\ and\ \bibinfo {author} {\bibfnamefont {O.}~\bibnamefont
  {Sarbach}},\ }\bibfield  {title} {\bibinfo {title} {Existence of the
  {C}hapman-{E}nskog solution and its relation with first-order dissipative
  fluid theories},\ }\href {https://doi.org/10.1515/jnet-2024-0086} {\bibfield
  {journal} {\bibinfo  {journal} {Journal of Non-Equilibrium Thermodynamics}\
  }\textbf {\bibinfo {volume} {50}},\ \bibinfo {pages} {295} (\bibinfo {year}
  {2025})}\BibitemShut {NoStop}%
\bibitem [{\citenamefont {Israel}(1963)}]{wI63}%
  \BibitemOpen
  \bibfield  {author} {\bibinfo {author} {\bibfnamefont {W.}~\bibnamefont
  {Israel}},\ }\bibfield  {title} {\bibinfo {title} {Relativistic kinetic
  theory of a simple gas},\ }\href@noop {} {\bibfield  {journal} {\bibinfo
  {journal} {J. Math. Phys.}\ }\textbf {\bibinfo {volume} {4}},\ \bibinfo
  {pages} {1163} (\bibinfo {year} {1963})}\BibitemShut {NoStop}%
\bibitem [{\citenamefont {Taylor}(1996)}]{Taylor96c}%
  \BibitemOpen
  \bibfield  {author} {\bibinfo {author} {\bibfnamefont {M.}~\bibnamefont
  {Taylor}},\ }\href {https://doi.org/10.1007/978-1-4419-7049-7} {\emph
  {\bibinfo {title} {Partial Differential Equations III: Nonlinear
  Equations}}},\ \bibinfo {edition} {2nd}\ ed.,\ \bibinfo {series} {Applied
  Mathematical Sciences}, Vol.\ \bibinfo {volume} {117}\ (\bibinfo  {publisher}
  {Springer},\ \bibinfo {address} {New York},\ \bibinfo {year}
  {1996})\BibitemShut {NoStop}%
\bibitem [{\citenamefont {Sarbach}\ \emph {et~al.}(2019)\citenamefont
  {Sarbach}, \citenamefont {Barausse},\ and\ \citenamefont
  {Preciado-L\'opez}}]{oSeBjP19}%
  \BibitemOpen
  \bibfield  {author} {\bibinfo {author} {\bibfnamefont {O.}~\bibnamefont
  {Sarbach}}, \bibinfo {author} {\bibfnamefont {E.}~\bibnamefont {Barausse}},\
  and\ \bibinfo {author} {\bibfnamefont {J.}~\bibnamefont {Preciado-L\'opez}},\
  }\bibfield  {title} {\bibinfo {title} {{Well-posed Cauchy formulation for
  Einstein-\ae{}ther theory}},\ }\href
  {https://doi.org/10.1088/1361-6382/ab2e13} {\bibfield  {journal} {\bibinfo
  {journal} {Class. Quant. Grav.}\ }\textbf {\bibinfo {volume} {36}},\ \bibinfo
  {pages} {165007} (\bibinfo {year} {2019})}\BibitemShut {NoStop}%
\bibitem [{\citenamefont {Kato}(1980)}]{Kato-Book}%
  \BibitemOpen
  \bibfield  {author} {\bibinfo {author} {\bibfnamefont {T.}~\bibnamefont
  {Kato}},\ }\href@noop {} {\emph {\bibinfo {title} {Perturbation Theory for
  Linear Operators}}}\ (\bibinfo  {publisher} {Springer-Verlag},\ \bibinfo
  {address} {New York},\ \bibinfo {year} {1980})\BibitemShut {NoStop}%
\bibitem [{\citenamefont {Israel}\ and\ \citenamefont
  {Stewart}(1979)}]{Israel1979}%
  \BibitemOpen
  \bibfield  {author} {\bibinfo {author} {\bibfnamefont {W.}~\bibnamefont
  {Israel}}\ and\ \bibinfo {author} {\bibfnamefont {J.}~\bibnamefont
  {Stewart}},\ }\bibfield  {title} {\bibinfo {title} {Transient relativistic
  thermodynamics and kinetic theory},\ }\href
  {https://doi.org/https://doi.org/10.1016/0003-4916(79)90130-1} {\bibfield
  {journal} {\bibinfo  {journal} {Annals of Physics}\ }\textbf {\bibinfo
  {volume} {118}},\ \bibinfo {pages} {341} (\bibinfo {year}
  {1979})}\BibitemShut {NoStop}%
\bibitem [{\citenamefont {Israel}(1989)}]{Israel1987}%
  \BibitemOpen
  \bibfield  {author} {\bibinfo {author} {\bibfnamefont {W.}~\bibnamefont
  {Israel}},\ }\bibfield  {title} {\bibinfo {title} {Covariant fluid mechanics
  and thermodynamics: An introduction},\ }in\ \href@noop {} {\emph {\bibinfo
  {booktitle} {Relativistic Fluid Dynamics}}},\ \bibinfo {editor} {edited by\
  \bibinfo {editor} {\bibfnamefont {A.~M.}\ \bibnamefont {Anile}}\ and\
  \bibinfo {editor} {\bibfnamefont {Y.}~\bibnamefont {Choquet-Bruhat}}}\
  (\bibinfo  {publisher} {Springer Berlin Heidelberg},\ \bibinfo {address}
  {Berlin, Heidelberg},\ \bibinfo {year} {1989})\ pp.\ \bibinfo {pages}
  {152--210}\BibitemShut {NoStop}%
\end{thebibliography}%

\end{document}